\documentclass[12pt,preprint]{aastex}

\def\undertext#1{$\underline{\smash{\hbox{#1}}}$}
\newcommand\diamondrule{\line{$\m@th
   \leaders\hrule\hfill\rlap{$\m@th\bracerd\braceld$}
   \braceru\bracelu\leaders\hrule\hfill$}}

\newcommand\upskip{\vskip-9pt}

\newcommand\bupskip{\vskip-17pt}
\newcommand\smallneg{\kern-.0800em}
\newcommand\negskip{\kern-.5em}

\newcommand\lsim{\rlap{\raise.4ex\hbox{$<$}}\lower.55ex\hbox{$\sim$}\,}
\newcommand\gsim{\rlap{\raise.4ex\hbox{$>$}}\lower.55ex\hbox{$\sim$}\,}
\newcommand\implies{{\bf=\kern-0.45em>}}

\newcommand\unit{\,\rm}

\newcommand\kms{\rm\, km\cdot s^{-1}}

\newcommand\um{\unit\mu m}

\newcommand\CO{\rm {}^{12}\smallneg CO}

\newcommand\COit{{}^{12}\smallneg\it CO}
\newcommand\cO{\rm {}^{13}\smallneg CO}

\newcommand\cObf{\bf {}^{13}\smallneg CO}

\newcommand\cOit{{}^{13}\smallneg\it CO}
\newcommand\Co{\rm C{}^{18}\smallneg O}

\newcommand\Jtwo{\rm J=2\rightarrow 1}

\newcommand\Jone{\rm J=1\rightarrow 0}
\newcommand\Jonebf{\bf J=1\rightarrow 0}

\newcommand\COone{\CO\ \Jone}

\newcommand\cOtwo{\cO\ \Jtwo}

\newcommand\Coone{\Co\ \Jone}
\newcommand\cOone{\cO\ \Jone}
\newcommand\cOonebf{\cObf\ \Jonebf}

\newcommand\NH{\rm N(H_2)}
\newcommand\nH{\rm n(H_2)}
\newcommand\nHbf{\bf n(H_2)}
\newcommand\nh{\rm n_{_{H2}}}
\newcommand\mh{\rm m_{_{H2}}}
\newcommand\ngr{\rm n_{{}_{gr}}}
\newcommand\sgr{\rm \sigma_{{}_{gr}}}
\newcommand\mgr{\rm m_{{}_{gr}}}
\newcommand\dgr{\rm \rho_{{}_{gr}}}
\newcommand\agr{{\hbox{$a$}}}
\newcommand\agi{{\hbox{$a_{_{min}}$}}}
\newcommand\aga{{\hbox{$a_{_{max}}$}}}
\newcommand\ai{{\hbox{$a_{_{min}}^{0.5}$}}}
\newcommand\ax{{\hbox{$a_{_{max}}^{0.5}$}}}
\newcommand\aef{{\hbox{$a_{_{eff}}$}}}
\newcommand\acc{\bar{\alpha}_{_T}}	
\newcommand\Tk{\rm T_{{}_K}}

\newcommand\Tkone{\rm T_{{}_{K1}}}

\newcommand\Tr{\rm T_{{}_R}}

\newcommand\ckms{\unit cm^{-2}\cdot (km\cdot s^{-1})^{-1}} 
\newcommand\ctkms{\ {}^{13}\smallneg CO\, molecules\cdot\unit cm^{-2}\cdot (km\cdot s^{-1})^{-1}}

\newcommand\Td{\rm T_{d}}
\newcommand\Tdz{\rm T_{d0}}
\newcommand\Tdo{\rm T_{d1}}
\newcommand\Tdc{\rm T_{dc}}
\newcommand\DT{\rm \Delta T}
\newcommand\rd{\rm r_{_{240}}}
\newcommand\rdl{\rm r_{_{1300}}}
\newcommand\Ia{\rm I_\nu(140\um)}
\newcommand\Ib{\rm I_\nu(240\um)}
\newcommand\Ic{\rm I(\cO)}

\newcommand\MJsr{\unit MJy\cdot sr^{-1}}
\newcommand\MJkk{\unit MJy\cdot sr^{-1}\cdot (K\cdot km\cdot s^{-1})^{-1}}

\newcommand\Ngas{\rm N(H\,I+2H_2)}

\newcommand\amrat{\rm N(H\,I)/2N(H_2)}

\newcommand\Dvc{\rm\Delta v_c}
\newcommand\nvco{\rm {N_{c1}\over\Dvc}}
\newcommand\nvcz{\rm {N_{c0}\over\Dvc}}
\newcommand\nvtco{\rm {N_{c1}(\cO)\over\Dvc}}
\newcommand\nvtcz{\rm {N_{c0}(\cO)\over\Dvc}}

\newcommand\NDv{\rm N(\cO)/\Delta v}
\newcommand\NDvbf{\bf N(\cO)/\Delta v}
\newcommand\Ndv{\rm {N(\cO)\over \Delta v}}
\newcommand\NDV{\rm N(\CO)/\Delta v}
\newcommand\Xdvdr{\rm X(\cO)/(dv/dr)}
\newcommand\XDvdr{\rm X(\CO)/(dv/dr)}

\newcommand\DGC{\rm\Lambda_{gd}}

\newcommand\degree{\rlap{$^\circ$}\kern.06em} 

\newcommand{\mymail}{wwall@inaoep.mx}

\slugcomment{Version -- 26/Mar/2007}

\shorttitle{Diagnostic of Dust/Molecular Gas Conditions}
\shortauthors{Wall}

\begin{document}


\title{Comparison of $\cObf$ Line and Far-Infrared Continuum\\
	Emission as a Diagnostic of Dust and Molecular Gas\\
	Physical Conditions:\\
	III. Systematic Effects and Scientific Implications}


\author{W. F. Wall}
\affil{Instituto Nacional de Astrof\'{\i}sica, \'Optica, y Electr\'onica,
Apdo. Postal 51 y 216, Puebla, Pue., M\'exico}
\email{\mymail}







\begin{abstract}

Far-infrared continuum data from the {\it COBE}/{\it DIRBE} instrument were combined with Nagoya 4-m 
$\cOone$ spectral line data to infer the multiparsec-scale physical conditions in the Orion$\,$A
and B molecular clouds, using 140$\um$/240$\um$ dust color temperatures and the 
240$\um$/$\cOone$ intensity ratios. In theory, the ratio of far-IR, submillimeter, or millimeter 
continuum to that of a $\cO$ (or $\Co$) rotational line can place reliable upper limits on the 
temperature of the dust and molecular gas on multi-parsec scales; on such scales, both the line
and continuum emission are optically thin, resulting in a continuum-to-line ratio that suffers no 
loss of temperature sensitivity in the high-temperature limit as occurs for ratios of CO rotational 
lines or ratios of continuum emission in different wavelength bands.  

Two-component models fit the Orion data best, where one has a fixed-temperature and the
other has a spatially varying temperature.  The former represents gas and dust towards the surface
of the clouds that are heated primarily by a very large-scale (i.e. $\sim 1\,$kpc) interstellar
radiation field.  The latter represents gas and dust at greater depths into the clouds and are
shielded from this interstellar radiation field and heated by local stars.  The inferred
physical conditions are consistent with those determined from previously observed maps of 
$\COone$ and $\Jtwo$ that cover the entire Orion$\,$A and B molecular clouds.  The models
require that the dust-gas temperature difference is 0$\pm 2\,$K.  If this surprising result applies 
to much of the Galactic ISM, except in unusual regions such as the Galactic Center, then there are 
a number implications.  These include dust-gas thermal coupling that is commonly factors of 5 to 10 
stronger than previously believed, Galactic-scale molecular gas temperatures closer to 20$\,$K than 
to 10$\,$K, an improved explanation for the N(H$_2$)/I(CO) conversion factor (a full discussion of
this is deferred to a later paper), and ruling out at least one dust grain alignment mechanism.  The 
simplest interpretation of the models suggests that about 40--50\% of the Orion clouds are in the form 
of cold (i.e. $\sim 3$-10$\,$K) dust and gas, although alternative explanations are not ruled out. 
These alternatives include the contribution to the 240$\um$ continuum by dust 
associated with atomic hydrogen and reduced $\cO$ abundance towards the clouds' edges.  Even
considering these alternatives, it is still likely that cold material with temperatures of 
$\sim 7$-10$\,$K still exists.  If this cold gas and dust are common in the Galaxy, then 
mass estimates of the Galactic ISM must be revised upwards by up to 60\%.

The feasibility of submillimeter or millimeter continuum to $\cO$ line ratios constraining 
estimates of dust and molecular gas temperatures was tested.  The model fits allowed the simulation 
of the necessary millimeter-continuum and $\cOone$ maps used in the test.  In certain ``hot spots" 
--- that have continuum-to-line ratios above some threshold value --- the millimeter continuum to 
$\cO$ ratio can estimate the dust temperature to within a factor of 2 over large ranges of physical 
conditions.  Nevertheless, supplemental observations of the $\cOtwo$ line or of shorter wavelength 
continuum are advisable in placing lower limits on the estimated temperature.  Even without such 
supplemental observations, this test shows that the continuum-to-line ratio places reliable upper 
limits on the temperature.  

\end{abstract}




\section{Introduction\label{sec1}}

While interesting in themselves, molecular clouds provide insights into star formation.  
Since stars form in and from molecular clouds, knowing the physical conditions within 
these clouds is essential for a complete understanding of star formation.  As mentioned
in Paper~I \citep{W05}, the warm (i.e., $\gsim 50$--100$\,$K) molecular gas associated
with star formation is often identified and diagnosed from observations of different
rotational lines of CO \citep[e.g.,][]{Wilson01, Plume00, Howe93, Graf93, Graf90, 
Boreiko89, Fixsen99, Harris85, Harr99, W91, Gus93, Wild92, Harris91}.   Molecular gas, 
and the interstellar medium in general, can also be observed in the millimeter, 
submillimeter, and far-IR continuum, which trace the emission of the dust grains 
associated with interstellar gas.  Continuum surveys can probe the structure 
and excitation of the ISM \citep[see, for example,][]{Dupac01, W96, Bally91, Zhang89,
Werner76, Heiles00, Reach98, Boulanger98, Lagache98, Goldsmith97, Sodroski94,
Boulanger90, Sellgren90, Scoville89, Sodroski89, Leisawitz88}.  Estimating physical
parameters like temperature, and sometimes density, requires using the ratios of
intensities of spectral lines or of the continuum at different wavelengths.  Given
that each of these ratios is dependent on the ratio of two Planck functions at two
different wavelengths, they often lose temperature sensitivity at higher temperatures. 
While there are methods of addressing this shortcoming, having two tracers of molecular
gas with different dependences on the temperature would complement other methods of
tracing warm dust or molecular gas.  This is especially true if the tracers are 
optically thin, because low opacity emission is more sensitive to the physical 
parameters of the bulk of the gas, rather than in just the surface layers.

One such pair of tracers is a rotational line of an isotopologue of CO, such as that of 
$\cO$ or $\Co$, and the submillimeter continuum.  Both of these tracers are optically
thin on the scales of many parsecs, which are the scales of interest for the current
work.  \citet{Schloerb87} and \citet{Swartz89} showed
that the intensity ratio of an optically thin isotopic CO line emission to submillimeter 
continuum emission can estimate the temperature of gas and dust in molecular clouds.
The \citet{Schloerb87} expression for this ratio goes roughly like T$^2$ in the 
high-temperature limit.  Accordingly, the $\rm I_\nu(submm)/I(\Coone)$ and 
$\rm I_\nu(submm)/I(\cOone)$ ratios are actually {\it more sensitive to temperature as 
that temperature increases.\/}  This is in stark contrast to ratios of rotational lines 
of a given isotopologue of CO and to ratios of continuum intensities at different 
frequencies, which {\it lose\/} sensitivity to temperature in the high-temperature limit. 
The $\rm I_\nu(submm)/I(\cO)$ ratio can then serve as the needed diagnostic of high
gas/dust temperatures, provided that the shortcomings and complications of these
tracers can be overcome or at least mitigated.  These complications include variations
in the $\cO$-to-dust mass ratio, non-molecular phases of the ISM along the line of
sight, variations of gas density, variations in dust grain properties, appreciable
optical depth variations in the $\cO$ line used, and others (see the Introduction of
Paper~I for more details).  Such complications are often reduced in the case of 
observations on multi-parsec scales, because spatial gradients on such scales are 
generally smaller than the extremes that occur on very small scales.  Consequently,
testing the reliability of the $\rm I_\nu(submm)/I(\cO)$ ratio as high-temperature
diagnostic is best carried out with observations of a molecular cloud, or of clouds,
on multi-parsec scales. 

The Orion~A and B molecular clouds were chosen as the clouds for testing the\hfil\break
$\rm I_\nu(submm)/I(\cO)$ ratio's diagnostic ability.  They have been mapped in the 
$\cOone$ line \citep[see, e.g.,][]{Nagahama98} and in the far-IR by {\it IRAS\/} 
\citep{Bally91} and {\it COBE/DIRBE\/} \citep[][W96 hereafter]{W96}.  Avoiding the
complication of the emission of stochastically heated dust grains requires far-IR
observations at wavelengths longer than 100$\um$ \citep[e.g.][W96]{Desert90}. 
Accordingly, the far-IR observations of {\it COBE/DIRBE\/} were used instead of
{\it IRAS\/} because the former has two bands --- $\lambda = 140\um$ and 240$\um$ ---
longward of 100$\um$, whereas the latter does not.  The Orion~A and B clouds were
chosen for this study because they have the advantages that they are bright in
$\cOone$ and at far-IR wavelengths, are out of the Galactic plane to avoid confusion 
with foreground and background emission, are several degrees in size so as to accommodate
many {\it DIRBE\/} beams, and have the best range of dust temperatures at the {\it DIRBE\/}
resolution of $0\degree.7$ \citep[see the Introduction of Paper~I and][for more details]{dirbex}.  
Therefore, the $\Ib/\Ic$ ratio, hereafter called $\rd$, was plotted against the 140$\um$/240$\um$ 
dust color temperature, or $\Tdc$, to test the $\rd$'s ability to recover molecular cloud 
physical conditions.  Physical models were applied to these data and physical conditions
were inferred in Paper~I.  The reliability of the model results were tested with 
simulated data in Paper~II \citep{W05a}.

The next section summarizes the model results (i.e. Paper~I) and the results of the 
simulations (i.e. Paper~II).   Section~\ref{ssec38} then discusses general systematic 
effects that had not been treated previously.  Section~\ref{sec4} gives the scientific
implications of the results.

\section{Review of the Results of the Modelling and of the Simulations\label{sec2}} 

The details of the treatment of the data and of the modeling and its results are found 
in Paper~I.  After a subtraction of large-scale emission from the Orion 140$\um$ and
240$\um$ maps representing foreground/background emission not associated with the 
Orion clouds (such subtraction was not necessary and, therefore, not applied to the
$\cOone$ map), one-component and two-component model curves were fitted to the 
observational data in the $\rd$ versus $\Tdc$ plot.  There were two types of 
one-component models: LTE and LVG (a type of non-LTE model).   There were also two
types of two-component models (both types being using the LVG code): simple 
two-component models and two-subsample, two-component models.  These models all
adopted some form of the following assumption:
\smallskip
\hfil\break
{\it The only physical parameters that change from one line of sight
to the next are the dust temperature, $T_{_d}$, and the gas kinetic 
temperature, $T_{_K}$, while maintaining a constant difference,
$\DT \equiv T_{_d} - T_{_K}$. Other physical parameters such as
gas density, dust-to-gas mass ratio, dust mass absorption coefficient,
etcetera are assumed to be constant from position to position.\/}
\smallskip
\hfil\break
This is referred to as the {\it basic assumption\/}.  In the case of the one-component,
LTE models, this means that the only fitted parameter was $\DT$, while the $\Td$ and
$\Tk$ freely varied from position to position.  For the one-component, LVG models, the
fitted parameters were $\DT$, the $\cO$ column density per velocity interval, $\NDv$, 
and the molecular hydrogen density, $\nH$, while the $\Td$ and $\Tk$ freely varied from 
position to position.  Put very explicitly, the basic assumption applied to the 
one-component, LVG models means that the $\DT$, $\NDv$, and $\nH$ were assumed to be
spatially unchanging and therefore are the parameters to be determined from the model
fits.  For the simple two-component models, there was a component~0, representing
dust and gas in the surface of the clouds and largely heated by a large-scale 
interstellar radiation field (ISRF), and a component~1, representing dust and gas deeper 
into the clouds heated by local stars and a large-scale ISRF attenuated by the surface 
layers of gas and dust.  The physical parameters of component~0 were spatially unchanging 
and the physical parameters of component~1 were also spatially unchanging, except for 
$\Td$ and $\Tk$.  The component-0 parameters were the dust temperature, $\Tdz$, the column 
density per velocity interval, $\nvtcz$, the density, $n_{c0}$, and the filling factor 
relative to component~1, $c_0$.  The component-1 parameters were $\nvtco$ and $n_{c1}$. 
(The component-1 dust temperature, $\Tdo$, varied so as to generate a curve in the $\rd$
versus $\Tdc$ plot.  Component 0, for example, would only generate a single point in
this plot if there were no component~1 contributing to the model output.) One more
parameter derived from the model fit was the $\DT$, which was assumed to be the same
for both components.  The two-subsample, two-component models were similar to the
simple two-component models, except that the models were fitted to two separate
subsamples within the sample of data points: the points with $\Tdc<20\,$K and those
with $\Tdc\ge 20\,$K.  Having two subsamples allowed a better fit to the $\Tdc\ge 20\,$K
points; the fits are normally dominated by the $\Tdc < 20\,$K subsample of points, often
preventing good fits to the $\Tdc\ge 20\,$K points.  The resultant parameter values for
all the model fits are summarized in Table~\ref{tbl-1}.  (Notice that the two-component
model results for the $\Tdc\geq 20\,$K subsample are shown for the two-subsample,
two-component models, whereas in Table~2 of Paper~I the one-component model results
were shown for this subsample.)

To check the results, the systematics were tested.  This was done by applying scale
factors to the model curves that represented the effect of systematic uncertainties.  
These were the uncertainties most directly related to the comparison between the model
curves and the observational data: the calibration of the observed $\Ib/\Ic$ ratio,
the uncertainty in the dipole moment of CO, the uncertainty in the $\cO$ abundance,
and the uncertainty in the dust optical depth to total gas column density.  A very
rough uncertainty of 20\% was adopted for each of these uncertainties (see Paper~I for
details).  These uncertainties are independent and, when added in quadrature, give a 
total systematic uncertainty of 40\% for the ratio of the model curve to the observed 
data.  Therefore, the scale factors applied to the model curves ranged between 0.6 and 
1.4.  They were chosen to change in steps of 0.2; the scale factors used were 0.6, 0.8,
1.0, 1.2, and 1.4.  Each scale factor was multiplied by the model curve before each fit
for all the model types.  This gave a range of fitted values for each parameter and this
range represents the systematic uncertainty for the given parameter. 

For the two-component models an additional test was to slightly shift the starting 
search grid before running each fit.  This also gave a range of results that was
comparable to the range found from changing the scale factor. 

As a further check on the results, masses and beam filling factors were derived for
each of the model types.  Specifically, the gas-derived column densities were compared
against the dust-derived column densities as a self-consistency check: if the model 
curve acceptably fit the data in the $\rd$ versus $\Tdc$ plot, then there should be little
scatter in the column density versus column density plot.  As an additional check, the
beam filling factors should be physically meaningful; they should be $\leq 1$, given
that they are area filling factors.   

One flaw of the above-mentioned tests is that they cannot guarantee that the true 
values are within the ranges of results found from changing the scale factor or the
starting grid.  The method employed here could be biased to ranges of results that are
far from the correct values.   Consequently, a third series of tests was performed by 
fitting the one- and two-component models to simulated data (see Paper~II).  The results
of the modeling and the tests are listed in Table~2 of Paper~II.  (See Table~\ref{tbl-8}
of the current paper for an updated version of that table.)  

The most basic result of this modeling was that the two-component models fit the data
better than the one-component models at the 99.9\% confidence level, according to the
F-test.  The dust-gas temperature difference, $\DT$, was found to be zero to within 1 
or 2$\,$K.  The component-0 dust temperature, $\Tdz$, was found to be 18$\,$K (uncertainty
to be discussed in Section~\ref{ssec38}).  The other parameters, such as the column 
densities per velocity interval and volume densities for the two components, were much 
less certain.  This is understandable given that the $\cOone$ line emission is well
approximated by the optically thin, LTE limit for much of the gas of the Orion clouds.  
Consequently, only rough lower limits could be applied to the densities and rough upper 
limits to the column densities per velocity interval.  The lower limit on the column 
density per velocity interval of component~1 is simply the column density per velocity 
interval of the clouds on the scale of the {\it DIRBE\/} beam.  This lower limit for
component~0 is near the lower limit of the master search grid.  

One important consequence of the two-component models is that there is about 60\% more
mass than would be inferred from the simpler one-component models.  This extra material
is in the form of cold (i.e. $\sim 3$--10$\,$K) dust and gas.

\section{Considerations of Systematics\label{ssec38}}

In this section we consider the systematics that are less directly connected with
the comparison between models and data.  Specifically, we examine the effects of 
changing various assumptions, including the basic assumption itself, on the results;  
we see how the results change if we neglect to subtract the background/foreground 
emission from the data, the effects of the emission of the dust associated with the 
H$\,$I gas, the effects for different values of the spectral emissivity index, the 
effects of a spatially varying $\cO$ abundance, the effects of varying the column 
density per velocity interval or density, and how the signal-to-noise filtering has 
affected the results. 

\subsection{The Effects of No Background/Foreground Subtraction\label{sssec381}}

As stated in Section~2.1 of Paper~I, there are uncertainties in the subtraction of the 
large-scale emission (i.e., on the scale of the entire map shown in 
Figure~2 of Paper~I).   This large-scale emission was subtracted from the
140$\um$, 240$\um$, and H$\,$I maps.  Even though this uncertainty was estimated
to be 10\%, it is still a good idea to see how this subtraction
affects the model results.  This was done by repeating the model in the
LTE and LVG, one-component cases and in the LVG, two-component case (entire
subsample) for the data with{\it out\/} the background/foreground subtraction.
The results were roughly similar to scaling up the data or,
equivalently, scaling down the model curve.  As such, the model results in the
tested cases were roughly equivalent to those obtained for data
that did indeed have the subtraction of the large-scale emission with a scale
factor of about 0.9 applied to the model curve.  Consequently, {\it  
the systematic uncertainty in the observed data is much smaller than that of
the total adopted calibration uncertainty of 40\%.\/}  This therefore implies 
that any reasonable estimate of the uncertainty in determining the appropriate 
level of the large-scale emission to be subtracted will have an even smaller 
effect on the model results (equivalent to adjusting the scale factor by a few 
percent).  

\subsection{The Effects of Dust Associated with HI\label{sssec382}}

\begin{figure}
\epsscale{0.60}
\plotone{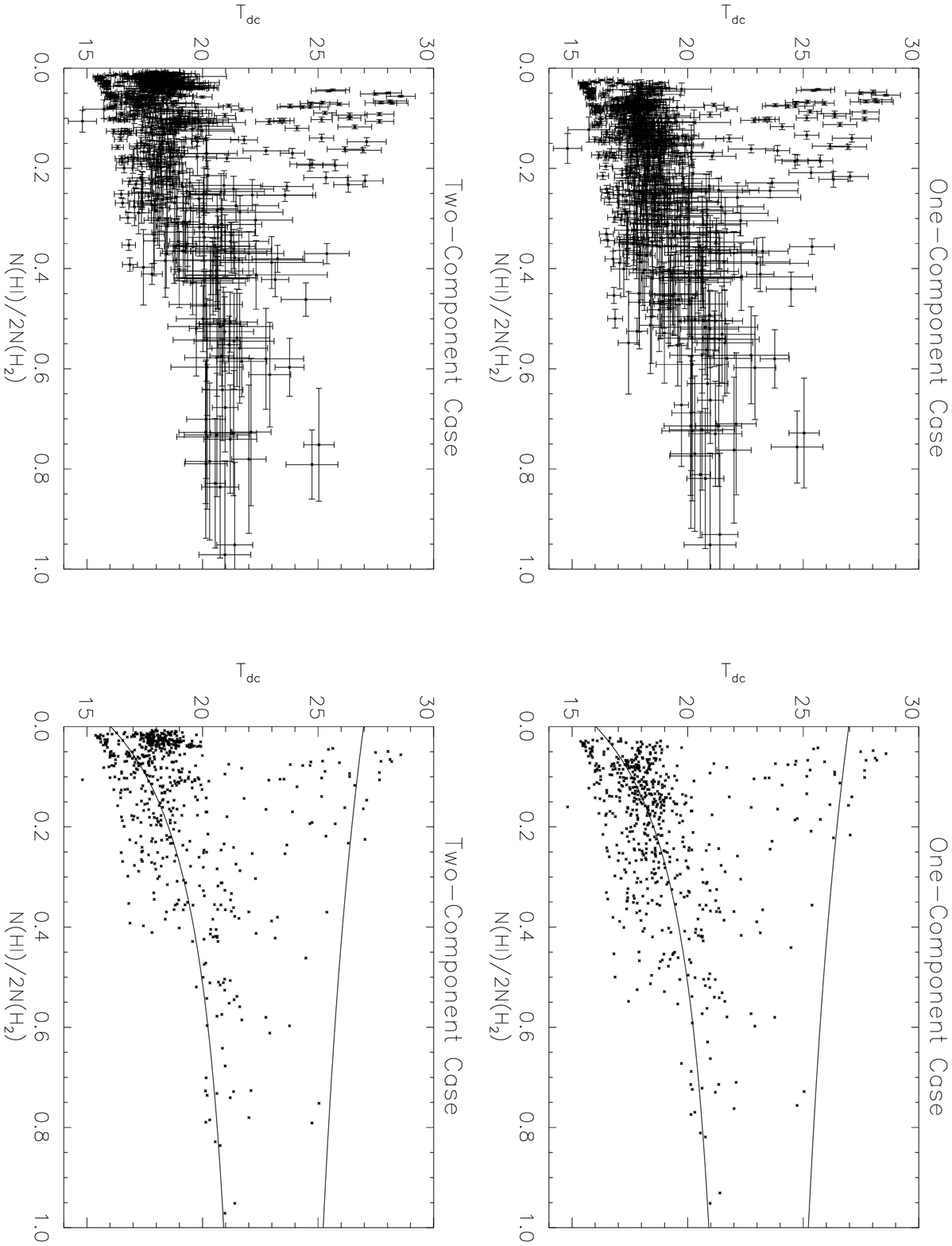}
\caption{Plots of the 140$\um$/240$\um$ dust color temperature versus the ratio
of the atomic gas to molecular gas column densities, $\amrat$, appear above.  
The upper two panels show these plots for the one-component, LVG models.  The
lower two panels are for the two-component, two-subsample, LVG models.  The 
panels on the left include the error bars, while the panels on the right exclude
the error bars.  The curves in the panels on the right represent hypothetical cases
where the dust associated with the molecular gas has one fixed temperature for all
lines of sight and the dust associated with the atomic gas has another fixed
temperature.  The lower curve in each of the right panels assumes that the dust
in the molecular gas has $\Td=16.5\,K$ and the dust in the atomic gas has 
$\Td=22.5\,K$.  The upper curve in each of the right panels assumes $\Td=27\,K$
and 22.5$\,$K for the dust associated with molecular and atomic gas, respectively.
The plots only include those pixels with the intensities above the 5-$\sigma$ level 
in $\Ia$, $\Ib$, $\Ic$  and above the 3-$\sigma$ level in I(H$\,$I) 
{\it simultaneously\/}. 
\label{fig55}}
\end{figure}

Even though the H$\,$I has only small column densities on all lines of sight
in the Orion fields (i.e. the average N(H$\,$I) is $5\times 10^{20}\unit cm^{-2}$ 
for those positions greater than 5-$\sigma$ in $\Ia$, $\Ib$, and $\Ic$), the dust 
associated with the H$\,$I may still have a non-negligible effect on the model 
results.  To test this, the 140$\um$/240$\um$ color temperature of the dust was 
plotted against the atomic gas to molecular gas column density ratio, $\amrat$.  
Figure~\ref{fig55} shows that the dust color temperature tends to about 
21$\,$K as the atomic to molecular gas ratio increases.  The correlation between 
the color temperature and the atomic-to-molecular gas ratio has a confidence level 
of better than 99.99\% according to the Spearman rank-order correlation test.  (More 
specifically, the significance of the null hypothesis of zero correlation is less 
than 10$^{-24}$.)  The curves represent the hypothetical case of having all the dust 
in the molecular gas at one fixed temperature and all the dust in the atomic gas at 
some other fixed temperature.  The lower curve assumes that the dust associated with 
the molecular gas has a temperature of 16.5$\,$K, while the upper curve assumes a 
dust temperature of 27$\,$K for the molecular-gas-associated dust.  Both curves 
assume that the atomic-gas-associated dust has a temperature of 22.5$\,$K.  Both 
curves together crudely describe the trends in the data.  Consequently, both curves
together imply that each line of sight either has molecular gas with cold dust, 
with $\Td=16.5\,$K, or with warmer dust, with $\Td=27\,$K, along with atomic gas 
that has dust with a constant temperature of 22.5$\,$K for every line of sight.  
The variations in dust color temperature would then be largely due to the variation
in the atomic-to-molecular gas ratio (along with some scatter).   This contradicts 
the picture represented by the models applied to the 
$\rd$ versus $\Tdc$ plot.  In that picture, the dust in the molecular gas does 
indeed vary in temperature from one line of sight to another, at least for the 
dominant component (i.e., component~1).  

\begin{figure}
\epsscale{0.8}
\plotone{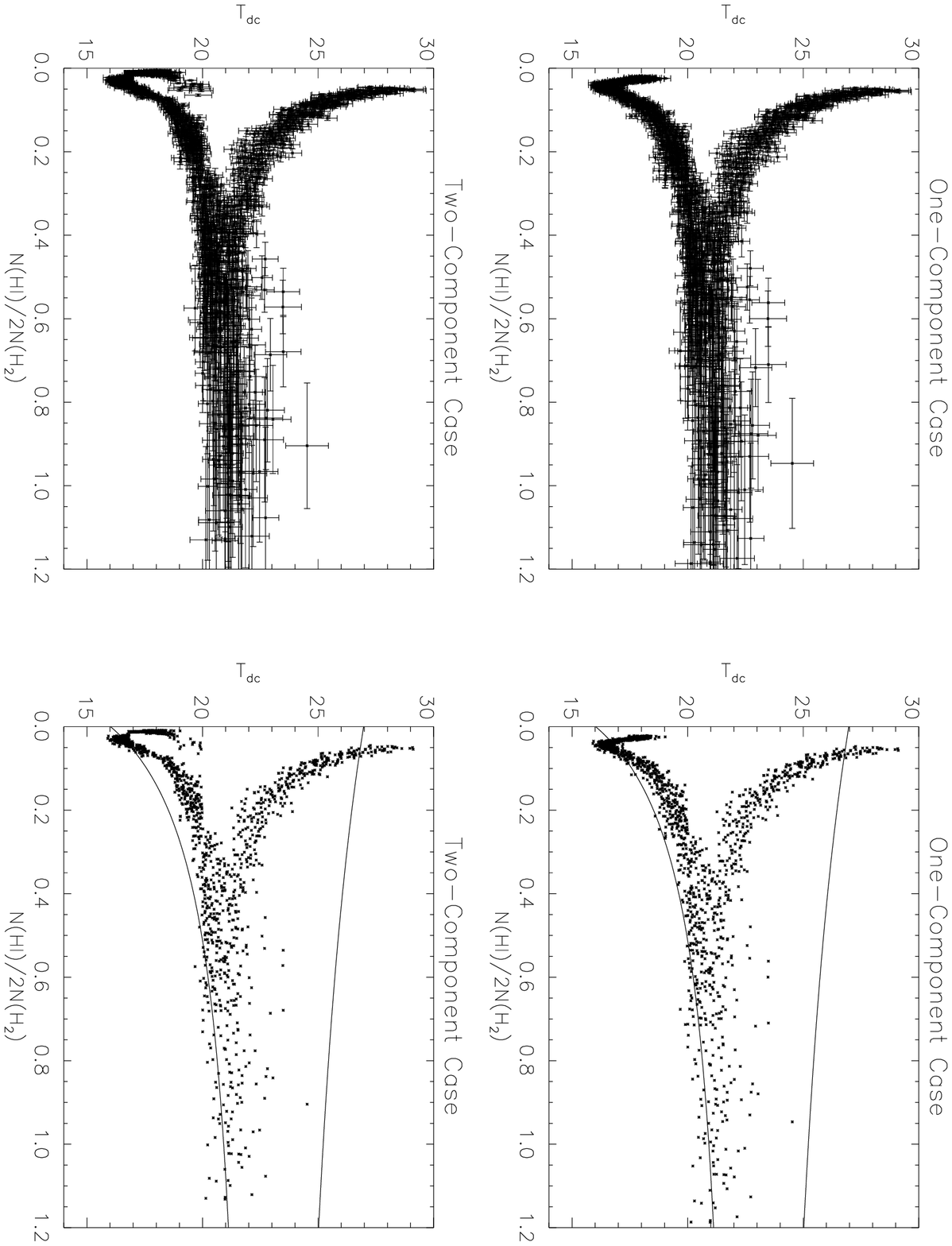}
\caption{The is the equivalent of Figure~\ref{fig55}, but for the 
{\it simulated\/} data.  The H$\,$I layer has a constant column density of
5$\times 10^{20}\, H\ atoms\cdot\rm cm^{-2}$ and a constant dust temperature 
of 22.5$\,$K.  The curves are the same as those in Figure~\ref{fig55}. 
\label{fig56}}
\end{figure}

To resolve this discrepancy, the simulated maps discussed previously
were modified by adding a layer of H$\,$I and its associated dust
with uniform properties throughout: a constant column density of 5$\times
10^{20}\, H\, atoms$$\unit cm^{-2}$ and a constant dust temperature of 22.5$\,$K.
Noise was added to the H$\,$I column density map that was the same as the 
value for the observed H$\,$I map.  The 140$\um$ and 240$\um$ intensities for
this H$\,$I layer were computed and added to the original noise-free maps.
The noise for the new continuum maps, that include the H$\,$I layer's dust 
emission, was then recomputed from the prescription used previously (i.e., 
expressions~1 and 2 of Paper~II).  The results are plotted in 
Figure~\ref{fig56} in the form of 140$\um$/240$\um$ color temperature 
versus atomic-to-molecular gas ratio, analogous to Figure~\ref{fig55}.
The simulated data does indeed reproduce the overall shape of the observed
data, despite the lack of scatter in the former compared to the latter.  
This scatter in the simulated data can be increased realistically by including 
variations in the H$\,$I column density map and in its dust temperature.
Nevertheless, it is clear that having {\it only\/} two possible constant 
dust temperatures in the dust associated with the molecular gas is {\it not\/} 
necessary for explaining the trends in Figure~\ref{fig55}.  Figure~\ref{fig56} 
shows the same overall trends even though it uses model results 
that allow a changing dust temperature in the molecular gas. 

The question remains as to the size of the change in the
model results because of this H$\,$I layer and its dust.  This question has
already been answered indirectly in the previous subsection, the subsection 
that dealt with the effects of no subtraction of the background/foreground 
(i.e., large-scale) emission.  Such a subtraction was necessary for the 
continuum maps (i.e., maps of dust emission) and for the H$\,$I map (i.e.,
map of atomic gas emission), but not for the $\cO$ map.  Therefore, the 
dust emission on the large scale is associated almost completely with the 
H$\,$I gas and {\it not\/} with the molecular gas.  Accordingly, {\it not\/}
subtracting the large-scale emission from the continuum maps is equivalent 
to piling on the atomic gas and its associated dust.  In fact, it is equivalent
to increasing the quantity of the {\it H$\,$I-associated\/} dust by factors of 
4 to 5 over that in the background/foreground-subtracted maps.  And yet, as
described previously, the effect of this extra dust ({\it not\/} associated
with molecular gas) was to change the results of the models (that {\it only\/} 
considered the dust associated with molecular gas) in a way consistent with
changing the scale factor by only about 10\%.  Therefore, the small amount 
of H$\,$I gas and its associated dust that remains in the 
background/foreground-subtracted maps will have an even smaller effect on the
model results --- equivalent to changing the scale factor by about 2 to 2.5\%.
It follows that the changes to the model results will be equivalent to only a 
tiny fraction of the full range of values for each parameter that is seen in 
such figures as Figure~21 of Paper~I or in Table~\ref{tbl-8}.  (Or in the case 
where there are only lower limits, such as those for the densities listed in 
Table~\ref{tbl-8}, those lower limits would be essentially unchanged.)

\subsubsection{The H$\,$I-Associated Dust and the Lower Temperature Limit
of the Cold Dust and Gas}

\begin{figure}
\epsscale{0.71}
\plotone{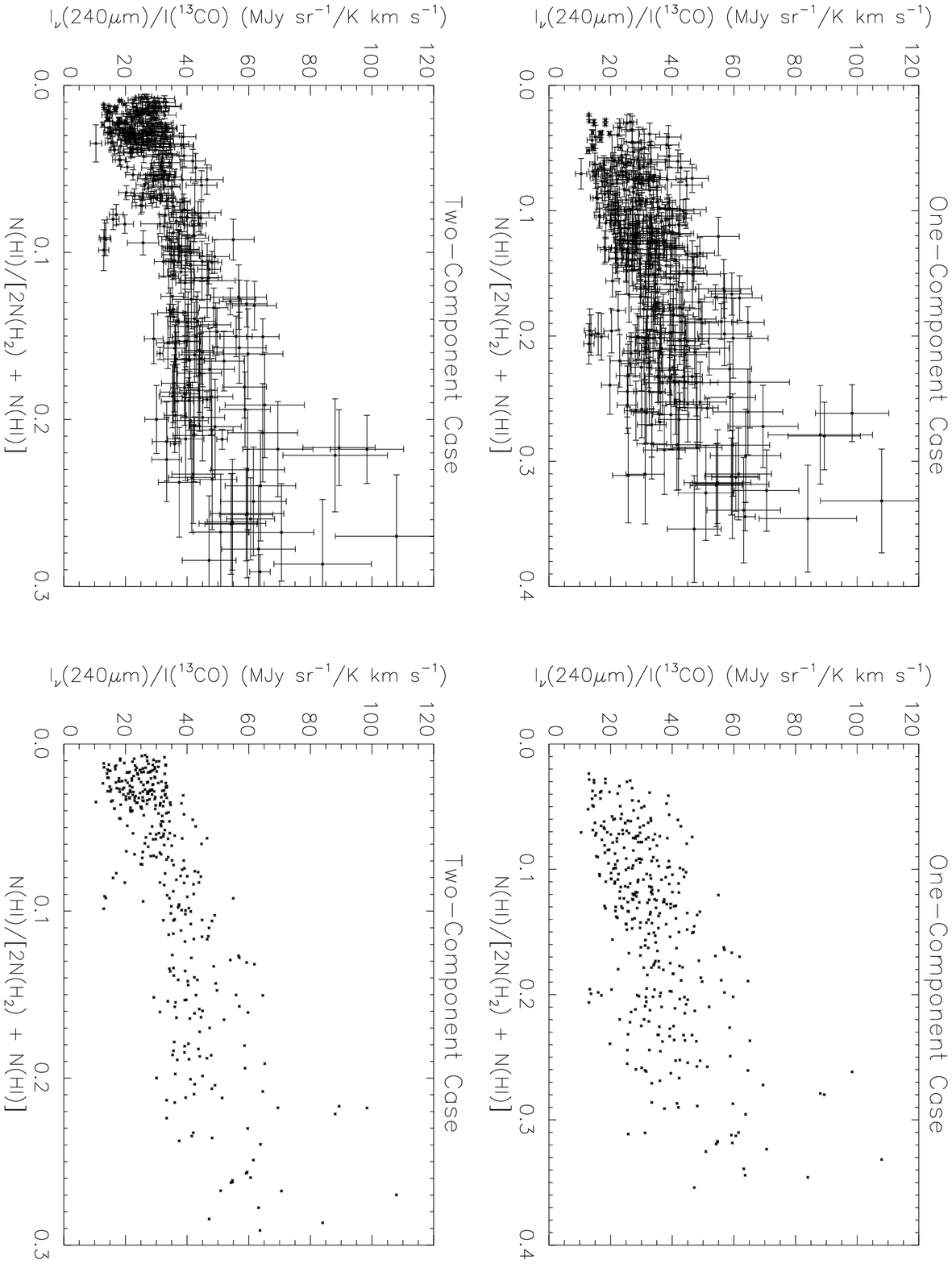}
\caption{Plots of the $\Ib/\Ic$ ratio versus the atomic hydrogen fraction,
N(H$\,$I)/[N(H$\,$I)+2N(H$_2$)] are shown for a subsample of the high signal-to-noise
positions with a 140$\um$/240$\um$ color temperature, $\Tdc$, near 18$\,$K.  Specifically, 
this sample of points is higher than 5$\sigma$ in $\Ia$, $\Ib$, and $\Ic$, higher than 
3$\sigma$ in I(H$\,$I 21$\,$cm), and with $\Tdc$ in the range 17 to 19.5$\,$K.  The
upper plots use the N(H$_2$) values of the non-LTE, one-component models and the lower
plots use those of the non-LTE, two-component, two-subsample models.  The left plots
include the error bars and right plots omit the error bars for clarity. 
\label{fig57}}
\end{figure}

One important point that remains is whether the dust associated with the H$\,$I is 
responsible for the model result that there is cold dust and gas at temperatures as 
low as 3$\,$K.  If so, then this result is incorrect.  As discussed before, the 
H$\,$I-associated dust has little {\it overall\/} effect on the majority 
of the plotted points.  However, this does not mean that the effect of this dust is 
negligible in {\it every sub-grouping\/} of points.  Specifically, the vertical section 
of the model curve in the $\rd$ versus $\Tdc$ plot (see Figures~20 and 24 of Paper~I) 
that represents this very cold material is located between $\rd\simeq
25$ and about 50$\MJkk$ for $\Tdc\simeq 18\,K$.  If the $\rd$ values were lower in 
this part of the plot, then no point would correspond to this
``cold" section of the model curve.  And the $\rd$ values would indeed be lower
if the effect of this H$\,$I-associated dust were removed, thereby allowing a higher 
lower limit on the gas and dust temperature of component~1 --- 
i.e., lower limits on $\Tkone$ and $\Tdo$ higher than 3$\,$K.  
Figure~\ref{fig57} shows plots of $\rd$ versus the H$\,$I fraction (i.e. 
N(H$\,$I)/[N(H$\,$I)+2N(H$_2$)]) for the positions with $\Tdc$ between 17 and 19.5$\,$K 
and with signal-to-noise ratios greater than or equal to 5 at 140$\um$, 240$\um$, in the 
$\cOone$ line, and greater than or equal to 3 in the H$\,$I 21-cm line.  These plots show 
a clear correlation (better than 99\% confidence from the Spearman rank-order correlation 
test), suggesting that the higher $\rd$ values are indeed due to dust associated with the 
atomic hydrogen.  The question is how strongly the H$\,$I-associated dust is contributing 
to the total 240$\um$ emission.  This allows us to correct for the emission 
of this dust, thereby effectively giving us only the continuum emission and line emission 
from the molecular gas alone.  

Estimating the appropriate correction to $\rd$ for the H$\,$I associated dust is far from 
straightforward.  One way is to make the crude assumption that the dust-emissivity per 
H-nucleon of the H$\,$I-associated dust, $\epsilon_{_{HI}}$, and for the H$_2$-associated
dust, $\epsilon_{_{H2}}$, are each constant for all the positions in the subsample of points 
to be tested.  This subsample is part of the sample used throughout this 
paper --- signal-to-noise greater than or equal to 5 for $\Ia$, $\Ib$, and 
$\Ic$ --- with the additional criteria mentioned above:
\begin{itemize}
\item[a)] Integrated intensity of the H$\,$I line greater than or equal to 3$\sigma$. 
\item[b)] $\Tdc$ between 17 and 19.5$\,$K.
\item[c)] $\rd$ between 25 and 50$\MJkk$.
\end{itemize}
The 240$\um$ specific intensity, $\Ib$, can be represented as
\begin{equation}
\Ib = \epsilon_{_{HI}}\, N(H\,I)\ +\ 2\epsilon_{_{H2}}\, N(H_2)\qquad .
\label{mr50}
\end{equation}
Simple linear regression in three dimensions can be used to solve for the emissivities,
$\epsilon_{_{HI}}$ and $\epsilon_{_{H2}}$, where we fit an equation of the form
\begin{equation}
z = ax\ +\ by\qquad ,
\label{mr51}
\end{equation}
solving for the optimal values of the coefficients $a$ and $b$.  Since normal linear regression 
only uses the uncertainties in the $z$ values, the fit should be repeated after interchanging the 
$z$ values with the $x$ values (and thereby using the $x$-uncertainties) and then 
again after interchanging the original $z$ values with the $y$ values (and thereby using the 
$y$-uncertainties).  This gives us three fits.  For the first fit, the
$\Ib$ are the $z$ and the uncertainties in $\Ib$ are used in the fitting, 
N(H$\,$I) is $x$, and 2N(H$_2$) is $y$.  The coefficients $a$ and $b$ then directly correspond 
to $\epsilon_{_{HI}}$ and $\epsilon_{_{H2}}$, respectively.  The second fit has N(H$\,$I) as $z$, the
uncertainties in N(H$\,$I) are used in the fitting, $\Ib$ is $x$, and 2N(H$_2$) is still $y$.
After solving for the optimum $a$ and $b$ values, expression~(\ref{mr51}) is rearranged to the form of
(\ref{mr50}) in order to solve for $\epsilon_{_{HI}}$ and $\epsilon_{_{H2}}$ in terms of $a$ and $b$.
The third fit is similar to the second fit, but with 2N(H$_2$) as $z$, N(H$\,$I) as $x$, and $\Ib$ as 
$y$.  Again, with the resultant $a$ and $b$ values, the expression~(\ref{mr51}) is rearranged to the
form of (\ref{mr50}) and $\epsilon_{_{HI}}$ and $\epsilon_{_{H2}}$ are found. (Another way to include 
the errors from all three quantities is to use the orthogonal regression method described at the 
beginning of Section~3 of Paper~I.)  In principle, the three different values determined for each of the 
emissivities can be used as a measure of their uncertainties.  

In practice, none of the fits were very good. The fits were applied using the 
molecular gas column densities (i.e., N(H$_2$)) derived from the one-component, non-LTE models and 
those using those column densities derived from the two-component, two-subsample, non-LTE models.  
Typical {\it reduced\/} chi-square values were from 30 to 200, although one was as low as $\chi_\nu^2
\simeq 6$.  None of the fits using the two-component model column densities had reduced chi-square 
values less than 80.  The poor quality of the fits reflects the invalidity of the assumption of spatially 
constant emissivities.  If we nonetheless use the one fit with the {\it least un}acceptable $\chi_\nu^2$ 
value of about 6, then $\epsilon_{_{H2}}\simeq 5\,\epsilon_{_{HI}}$.  Given that the maximum 
atomic gas fraction is about 0.3 (see Figure~\ref{fig57}), the H$\,$I-associated dust 
contributes less than about 7\% to the total 240$\um$ emission.  If we tighten criterion~c) above to $\rd$ 
between 40 and 50$\MJkk$ to see how the results are changed, and if we ignore the fits with $\chi_\nu^2
\gsim 10$ or with a negative value of either $\epsilon$, then we find that $\epsilon_{_{H2}}\simeq 9\, 
\epsilon_{_{HI}}$.  The least unacceptable fits suggest that the H$\,$I associated dust makes
a contribution of about 3 to 7\% to the 240$\um$ emission.  This result is similar, at least qualitatively,
to the result found previously that the H$\,$I-associated dust emission has a negligible effect on whole 
sample of high signal-to-noise points.

Given that no fit was acceptable, at least a simplistic correction 
should be applied to see what changes to the lower limit of the dust and gas temperature of component~1 
are {\it possible\/}.  The simplest kind of correction is to assume that the emissivities per 
H-nucleon are equal for the H$\,$I-associated dust and the H$_2$-associated dust, a much more extreme 
correction than the fits to equation~(\ref{mr50}) would suggest.  Accordingly, the observed $\Ib$ 
values, and therefore the $\rd$ ratio values, must be scaled by 2N(H$_2$)/[N(H$\,$I)+2N(H$_2$)].  
These correction factors were applied, using the one-component values for the molecular gas column 
densities.  This is obviously not consistent with the two-component model results that predict the
cold gas and dust.  However, assuming correction factors from the one-component model column 
densities results in more extreme (i.e. further from unity) values of the correction factors that permit 
a greater appreciation of the tight constraint on the lower temperature limit.  These correction 
factors were applied to the whole sample of high signal-to-noise positions --- specifically, all the 
points with $\Ia$, $\Ib$, and $\Ic$ greater than 5$\sigma$.  Since the H$\,$I data were used, criterion~a) 
reduced the sample to 609 points.  The uncertainties in 
the correction factors were {\it not\/} propagated to the error bars in the $\rd$ values in the sample.  
The larger error bars in the $\rd$ ratios would have produced an ambiguity in the interpretation of the 
results: was the change in the lower temperature limit of component~1 really due to the correction for 
the H$\,$I-associated dust or was this simply due to the larger error bars?  The results of fitting the 
two-component models to the corrected data should not be taken literally in any case and only point 
to the potential effects of correcting for the unwanted continuum emission.  We are testing the 
hypothetical case of observing imaginary molecular clouds absolutely free of atomic hydrogen and its
associated dust.  

The results of fitting a simple two-component model to the H$\,$I-corrected data are consistent with 
the previous two-component model results in Section~3.3 of Paper~I.  In particular, the results of
these corrected models are within the range of values depicted in Figure~21 of Paper~I
(ignoring the $c_o$ versus scale factor and the $\nvcz$ versus scale factor plots in favor of the
$c_o\nvcz$ product versus scale factor plot), except that the density $n_{c0}$ is as small as 
10$\unit cm^{-3}$.   One noteworthy difference is that $\chi_\nu^2$ is higher for the two-component 
model fitted to the corrected data, $\chi_\nu^2=6.41$, than for the model fitted to the uncorrected 
data, $\chi_\nu^2=5.69$.  If we now increase the lower temperature limit of component~1 from 
2.8$\,$K to 5$\,$K and search for the optimum parameter values again, then $\chi_\nu^2$ increases to 
7.54.  Specifically, we set the lower limits of $\Tkone$ to 5$\,$K and of $\DT$ to 0$\,$K, resulting 
in both $\Tkone$ and $\Tdo\geq 5\,$K.   Again keeping in mind that the points in the sample are not 
completely independent, the effective number of degrees of freedom is about 60.  Adopting this number, 
the F-test tells us that raising the lower limit of component-1 temperature to 5$\,$K can only be 
rejected at a confidence level of nearly 75\%.  Such a confidence level does not inspire much confidence, 
and we should probably accept the fit.  Consequently, it is {\it possible\/} that gas and dust with 
temperatures between 3 and 5$\,$K are not necessary for explaining the observations.  If this is the 
case, then the correction factor of 1.6 to the one-component masses would have to be corrected 
downward. 

In general, it {\it seems\/} that the H$\,$I-associated dust can{\it not\/} provide an alternative 
explanation for the cold dust and gas (i.e. temperature between 3 and 10$\,$K).  Nonetheless, there 
is a possibility that the lower limit of the component-1 temperature is about 5$\,$K instead of 3$\,$K.  
Then the estimated fraction of the total gas mass in the cold portion would be less than the original
estimate of about 40-50\%. 

\subsection{Varying X($\cObf$)\label{sssec383}}

\citet{Warin96} examined the photodissocation and rotational excitation of CO and its
isotopologues in diffuse, translucent, and dense dark clouds.  They found that the $\cO/\CO$
abundance ratio (i.e. X($\cO$)/X($\CO$)) can vary by factors of 2 or 3 as a function of
depth into the cloud (see the left-most panel of their Figure~16).  Given that $\Td$ 
varies with $\Ngas$ in the Orion clouds (e.g., see Fig~4 of Paper~II), is it then possible 
that the observed variation of $\rd$ with $\Td$ is because X($\cO$) is varying with $\Ngas$?  
Examining the I($\cO$)/I($\CO$) as a function of $\Ngas$ (an approximation of their Figure~16)
shows no consistent trend and is at odds with the results of \citet{Warin96}.  Therefore, it
is very {\it un}likely that variations of X($\cO$)/X($\CO$) could account for the {\it overall\/}
trends seen in $\rd$ versus $\Td$.  

Nevertheless, variations in X($\cO$)/X($\CO$) could still
explain unusually small or large $\rd$ values for some points in the $\rd$ versus $\Td$ plots. 
In particular, the points near the top of the triangular cluster could be explained
by a lower X($\cO$).  This is discussed in detail in Section~\ref{ssec44}.

\subsection{Models with $\bf\beta\ne 2.0$\label{sssec384}}

\citet{Johnstone99} find that the dust emissivity index, $\beta$, might have
a range as extreme as from 1.5 to 2.5 in Orion$\,$A.  Two-component models 
using adopted values of 1.5 and 2.5 for $\beta$ were applied to the data to see 
the effects on the results.  The resultant 
parameter values were consistent with the ranges of values listed in 
Table~\ref{tbl-8}, with one exception: the range of $\Tdz$ values is changed 
because $\beta$ is different from before --- all the inferred dust temperatures 
are changed.  For $\beta=1.5$, $\Tdz=21\,$K was found and, for $\beta=2.5$,
$\Tdz=16\,$K was the best fit.  

In short, with the exception noted above, the resultant parameter values are 
still within the range of values expected from the scale 
factor variations.  Thus the most extreme variations of $\beta$ still largely
give model results within the ranges listed in 
Table~\ref{tbl-8}.

\subsection{Models with Varying $\NDvbf$ or Varying $\nHbf$\label{sssec385}} 

These were essentially one-component models with only three parameters on
any {\it one given line of sight:\/} $\DT$, $\NDv$ and $\nH$.  The approach 
here was to vary not only the $\Td$ and $\Tk$ from one line of sight to 
another (keeping $\DT$ constant), but also to vary one of two parameters: 
the column density per velocity interval ($\NDv$) or the density ($\nH$). 
To characterize all the lines of sight represented in the $\rd$ versus
$\Td$ plot, there were 9 parameters in total for a curve that
was strongly decreasing through the triangular cluster of points at low
$\Td$, gently rising for the intermediate $\Td$, and then strongly rising 
for high $\Td$.  Thus there were 3 intervals of $\Td$ that were delimited
with 4 parameters: $T_1$, $T_2$, $T_3$, and $T_4$.  If we consider first
only the models with varying $\NDv$, then on the first interval, $[T_1,T_2]$,
the $\NDv$ was equal to $N_1$ at $T_1$ and to $N_2$ at $T_2$ and the $\NDv$
value at any point in this interval was determined by logarithmically 
interpolating between $N_1$ and $N_2$.  The same approach was used on the
last interval, $[T_3,T_4]$, with $\NDv$ values equal to $N_2$ at $T_3$ and 
to $N_3$ at $T_4$.  On the interval $[T_2,T_3]$, $\NDv$ was held constant
at a value of $N_2$.  The fitted parameters were then $\DT$, $T_1$,
$T_2$, $T_3$, $T_4$, $N_1$, $N_2$, $N_3$, and the fixed density, $n$
(with best-fit values of $0\pm 1\,$K, $16\pm 1\,$K, $17\pm 1\,$K, $24\pm 1\,$K, 
$27\pm 1\,$K, $(5\pm{3\atop 2})\times 10^{16}\ctkms$, $(2\pm{1\atop ?})\times 
10^{14}\ctkms$, $(8\pm 5)\times 10^{16}\ctkms$, and $(3\pm{2\atop 1})\times 
10^3 H_2\, molecules\cdot\unit cm^{-3}$, respectively).  Achieving a
reasonable-looking ``best" fit presented a number of problems:
\begin{enumerate}
\item The fitting is heavily biased towards the low-$\Td$ points (i.e.,
$\Td\lsim 21\,$K).  The points with $\Td\gsim 22\,$K are almost excluded
in the model fits. 
\item Even within the low-$\Td$ points, the fits tend to bypass the 
central area of the large triangular cluster of points.
\item There is a very large hump in the model curve for $\Td$ 
around 22$\,$K that overestimates the data points by more than an order
of magnitude. 
\item The best fit for $N_2$ is more than order of magnitude lower than 
the observed large-scale $\NDv$ value (see Section~3.2 of Paper~I).
\item In order to fix the bias towards the low-$\Td$ points and remove
the hump, it was necessary to apply weights to the data to reduce or remove 
this bias.  This resulted in a chi-square value that was somewhat subjective, 
given that the choice of weights was somewhat subjective.
\end{enumerate}

    The models with varying $\nH$ had similar, though not identical, problems.
For these models, the $\NDv$ parameters interchange with the $\nH$ parameters:
$\DT$, $T_1$, $T_2$, $T_3$, $T_4$, $n_1$, $n_2$, $n_3$, and the fixed column
density per velocity interval, $N$  (with best-fit values of $-2\pm 1\,$K, 
$14\pm 1\,$K, $18\pm 1\,$K, $22\pm 1\,$K, $30\pm 1\,$K, $10\pm{8\atop ?}\, H_2 
\, molecules\cdot\unit cm^{-3}$, $(5.6\pm{4.4\atop 2.5}\times 10^4\, H_2
\, molecules\cdot\unit cm^{-3}$, $10\pm{8\atop ?}\, H_2 \, molecules\cdot\unit 
cm^{-3}$, and $(2\pm{1\atop ?})\times 10^{15}\ctkms$, respectively).  As with
the models mentioned in the previous paragraph, these models have biases that 
inhibit good fits through the $\Td\gsim 22\,$K points or through the central 
part of the triangular cluster.  However, the problems for $\nH$-varying
models differ in three ways from those of the $\NDv$-varying models:
\begin{itemize}
\item The hump for the $\NDv$-varying models centered at about
$\Td=22\,$K is much smaller for the $\nH$-varying models.  This hump for
the latter models only overestimates the data by about a factor of 2.
\item The best-fit $\NDv$ value is still low, but is now within a factor
of 2 of the lower limit imposed by the large-scale observed $\NDv$ value.
\item The best way to reduce the bias towards the low-$\Td$ points was 
to fix the $T_1$, $T_2$, $T_3$, and $T_4$ values.  The choice of these 
values and their uncertainties was somewhat subjective. 
\end{itemize}
The best-fitting model curve gave a reduced chi-square of 9.2. 

    While the best of these models is much better than
the one-component models that used the basic assumption (i.e., $\chi_\nu^2
= 16.5$ and 16.9), it is still much worse than the two-component
models (i.e., $\chi_\nu^2 = 5.7$ and 5.3).  In addition, the models with
varying $\nH$ (or varying $\NDv$) require subjective judgement in determining
the values to fix for some of the parameters (or in adjusting the weights
of different subsamples of data points).  And despite having fewer parameters
(9 as opposed to 15) than the two-component, two-subsample models --- the most 
elaborate of the models that obey the basic assumption (within each subsample 
at least) --- the models with varying $\nH$ or varying $\NDv$ are actually 
{\it more\/} complicated than these two-component models.  They are more
complicated because these models essentially divide up the sample of data
points into {\it three\/} subsamples instead of two, and two of these
subsamples (in intervals $[T_1,T_2]$ and $[T_3,T_4]$) allow $\nH$ (or $\NDv$)
to vary in addition to $\Td$ (along with $\Tk$ so as to keep $\DT$ constant).
In contrast, the two-component, two-subsample models have only $\Td$ varying
(along with $\Tk$ so as to keep $\DT$ constant) within each subsample ---
the basic assumption is obeyed in each subsample. 

    In short, relaxing the basic assumption by using models with simple systematic 
variations of either $\NDv$ or of $\nH$ results in poorer, and more subjective, fits
than the best models following this assumption.

\subsection{Signal-to-Noise Considerations\label{sssec386}}

Given that the sample of points modeled in the $\rd$ versus $\Tdc$
plots represents only 25.8\% of the Orion Fields, how well does
this sample represent the physical conditions in the Orion
clouds as a whole?  A related question is what fraction of the positions
that have gas/dust emission from the Orion clouds is represented by the sample?  
The contour maps in Figure~2 of Paper~I suggest that roughly three-quarters of the 
Orion Fields have gas and dust.  Some low-level emission could come from 
elsewhere along the line of sight, so maybe only half of the Orion Fields 
are occupied by the Orion clouds.  However, even in this extreme possibility,
the modeled positions still only represent about 50\% of the Orion clouds' 
area --- still {\it not\/} a majority of the positions.  Clearly, just how 
strongly the inferred physical conditions depend on the signal-to-noise threshold 
used in sample selection must be tested. 

The most straightforward test, namely lowering the signal-to-noise 
threshold until most of the Orion Fields are represented, is 
not practical.  There would be 2 to 4 times the number of points to fit 
and the majority of these would have huge error bars.  Consequently, 
this sample of points would be difficult to model reliably.  The alternative
test is to raise the signal-to-noise threshold and see how the 
inferred physical conditions change with this threshold.  This has the 
disadvantage of going from a minority of the points to an even smaller 
minority.  Nonetheless, this alternative has the very strong advantage that
only very high signal-to-noise points are modeled, thereby yielding 
more reliable fit parameters. 

To do this alternative test, the threshold was increased appreciably: from 
5$\sigma$ to 20$\sigma$ in $\Ia$, $\Ib$, $\Ic$.  This sample has
only 6.6\% of the Orion Field positions.  An LVG, two-component model was fitted 
to this sample.  The resultant parameter values were consistent with those specified 
in Table~\ref{tbl-8}, with one notable exception: $\Tdz$ was 17$\,$K
instead of the usual 18$\,$K.  This change in the component-0 dust temperature
is not surprising.  The signal-to-noise ratio is proportional to the surface 
brightness, and this depends mostly on the gas/dust column density.  So 
signal-to-noise ratio is roughly equivalent to depth into the clouds.  Increasing
this ratio's threshold is almost like filtering out the cloud
edges and looking more deeply into the clouds.  The $\Tdz$ is the 
roughly constant temperature on the scale of the Orion clouds of some dust 
component.  If this dust component is on the surfaces of the 
clouds, then it is the temperature of the dust heated primarily by the 
general ISRF.  If this dust component is just below the clouds' surfaces,
then this is the temperature of the dust heated primarily by an ISRF shielded 
by the surface layers of gas and dust, resulting in a lower
temperature.  Then by extrapolation, modeling of {\it all\/} the positions
in the Orion~Fields (if there were sufficient signal-to-noise) could 
yield $\Tdz\simeq 19\,$K.  

Accordingly, modeling the full spread of points in the $\rd$ 
versus $\Tdc$ plot probably requires a spread in the physical parameter values 
that have been held constant.  For example, the horizontal spread of the triangular 
cluster in an $\rd$ versus $\Tdc$ plot (e.g., see Figure~20 of Paper~I)
probably means a component-0 dust temperature varying between about 16 and
19$\,$K. (As such, an updated version of Table~2 of Paper~II becomes Table~\ref{tbl-8}
of the current paper, which includes this estimate of the range of $\Tdz$ values.)  
Analogously, other parameter values, such as the densities and 
column densities per velocity interval, must also vary to ``fill'' the space 
occupied by the sample of points, as suggested in Section~3.5 of Paper~I.  Nonetheless,
the current modeling effort is a sufficient first approximation. 

The most important result of varying the signal-to-noise ratio is that the sample 
with only 25.8\% of the Orion~Field positions {\it may\/} indeed represent the 
bulk of the Orion clouds.  There was no appreciable change (except a slight
change in $\Tdz$) in going from the factor of $\sim$4 from 6.6\% of the fields
to 25.8\%.  Therefore, extrapolating the extra factor of $\sim$4 to 100\% {\it may\/} 
also yield no appreciable change.  This is far from 
certain, of course, because the extrapolation from 25.8\% to 100\% could
cross some depth threshold that appreciably alters the parameter values.
On the other hand, the effective spatial resolution of these observations
is 8$\,$pc at the distance of the Orion clouds.  
Averaging over such large size scales may reduce the effects of varying
the sample. 

In short, varying the signal-to-ratio used in selecting the sample
may have little effect on the derived gas/dust physical 
conditions.

\section{Scientific Implications and Discussion\label{sec4}}

The astronomical implications of the results are numerous and include
modifications to models of dust/gas thermal coupling, to estimates of mass and 
kinetic temperature of molecular gas on galactic scales, and to our understanding 
of the X-factor to name a few.  These issues and others will be discussed in the
following sections (or in subsequent papers as for the X-factor), after discussing 
the appropriateness of using the LVG models.

\subsection{Use of the LVG Code\label{ssec41}}

Some literature suggests that the LVG models can give
inconsistent results and that photodissociation region (PDR) models  
remove the inconsistencies \citep[e.g.,][]{Mao00}.  The PDR models 
represent a nearly complete explanation for the emission strength of molecular 
lines, whereas the LVG models merely relate the molecular line strength to simple 
physical parameters in a simplified case.  Consequently, the 
PDR models should yield more reliable estimates of physical parameters like 
temperature and density than those of the LVG models.  However, some papers that use 
CO line ratios to claim that the 
LVG model results are unsatisfactory compared with those of PDR models 
often suffer from flawed and inconsistent arguments.

\citet{Mao00}, for example, model the physical conditions in the central $\sim$500$\,$pc
of the galaxy M$\,$82.  They claim that their CO data yield physically unreasonable 
results when using the LVG models --- results that supposedly become more reasonable 
when using PDR models. 
The problems with this claim is summarized below. 
They state, among other things, that the LVG models imply gas densities that are too 
low, cloud sizes that are too large, and area filling factors that are inconsistent 
with volume filling factors.  Their first claim of low gas densities 
is connected with their restriction of the $\Xdvdr$ value.  They use the observed
{\it large-scale\/} velocity width of the lines compared with the size of the 
observed region to fix the numerical value of $\Xdvdr$ within a narrow range.
However, the large-scale (i.e., on the scale of the beam) $\Xdvdr$ has little to do 
with this parameter value within the individual clumps responsible 
for much of the observed emission.  If their approach {\it were\/} also applied to the 
volume density, then it {\it would have been\/} equivalent to dividing the mass of gas 
within the beam by the volume of this gas to estimate and fix the gas density, which is 
well known to be only a rough lower limit to the density 
within the clumps.  They did not make this mistake, but did make the equivalent mistake 
for the $\Xdvdr$ value.  Had they allowed $\Xdvdr$ to vary over a wider range, they 
would have satisfied the density constraints suggested by the other observations they 
mentioned.  Their second claim of overly large cloud sizes implied 
by the LVG model results was based on two different methods.  Both methods,
however, overestimated the cloud sizes by the ratio of the observed line velocity width 
to the cloud velocity width, which is about an order of magnitude.  One method 
was dividing the column density of a single cloud by its volume density, where both the 
column density and volume density were LVG results.  The second method estimated volume 
and area filling factors from LVG results, their ratio giving the size of a single cloud.  
Both methods were applied incorrectly.  The first method converted from 
$\NDv$ to cloud N(H$_2$) by using the {\it entire\/} observed line width, which
also includes the rotation of the observed galaxy, instead of  
using an estimate of a cloud line width (i.e. 5-10$\kms$).   The second method 
suffered from the same overestimate.  The cloud size depends on the 
ratio of the volume to area filling factors, but the area filling factor {\it must be\/} 
for {\it all\/} the gas at {\it all\/} the velocities within 
the line profile (assuming that there is never more than one cloud on any line of sight 
for {\it all\/} the velocities within the line profile).  However, they clearly 
used the ratio of the observed line strengths to those of the model, which gives the area 
filling factor {\it within a narrow velocity interval\/} and {\it not\/} the area filling 
factor over the entire line profile.  This second method must include 
the ratio of the cloud velocity width to the line velocity width, again 
reducing the estimated cloud size by an order of magnitude.  The third claim of 
inconsistent area and volume filling factors is weak at best, 
given that the relationship between the two is not a fixed, 
straightforward expression that applies in every case.  If we nonetheless accept the 
expression used by \citet{Mao00}, then there was indeed a minor discrepancy between the 
two types of filling factor for the LVG model results.  However, what \citet{Mao00} 
ignored entirely was that {\it the corresponding discrepancy for the PDR 
models is\/} {\bf much} {\it larger than that for the LVG model results.\/}  The filling 
factor argument was applied in a clearly biased manner.  In short, all three
claims are based on arguments that are faulty or biased or both.

In addition to the problems above, there is the strong evidence provided by
\citet{Weiss01}: they used data similar to those of \citet{Mao00} and 
recovered the same physical conditions, {\it 
while using only LVG models.\/}  Obviously, if \citet{Weiss01} and \citet{Mao00} agree
on the physical conditions, then they disagree on the necessity of the PDR models
for recovering those conditions.  And given that \citet{Weiss01} recovered those
conditions with only the LVG models, then the claim of \citet{Mao00}
that PDR models are necessary is clearly incorrect.  Therefore, the LVG 
models are {\it clearly as reliable\/} as the PDR models when using CO lines (at least 
on scales of a few or more parsecs). 

Consequently, the LVG models applied to the large-scale physical conditions in 
the Orion molecular clouds, as described in this paper, are also as reliable as 
the PDR models. 

\subsection{Comparison of Derived $\bf T_{_K}$, $\bf n(H_2)$, and $\bf N(\cObf)/\Delta v$ 
with Previous Work\label{ssec42}}

There are few papers that discuss the molecular gas physical conditions of the 
{\it entire\/} Orion$\,$A and B clouds, as inferred from two or more 
transitions of CO.  \citet{Sakamoto94} inferred these physical conditions using
the $\Jtwo$ and $\Jone$ lines of $\CO$.  Of all the rotational lines of $\CO$ and 
its isotopologues, these are the {\it least\/} sensitive to 
the gas physical conditions.  Nevertheless, they at least provide a rough 
comparison with the physical conditions obtained in the current paper that used 
the dust-continuum to gas-line ratio.   As discussed in Paper~II, the 
$\cOone$ emission is dominated by that of component~1 in all but the few points 
with $\Td=\Tk\lsim 4\,K$ in that component. Since this is molecular gas emission, 
this emission, rather than the FIR continuum, better identifies the component with 
which to compare single-component model results, such as those of \citet{Sakamoto94}.  
Consequently, the component-1 parameters in Table~\ref{tbl-8} are compared with the
\citet{Sakamoto94} results. 

The physical conditions inferred by \citet{Sakamoto94} are consistent with those 
inferred in the current paper.  They found the following physical conditions in the 
Orion molecular clouds: 
\begin{itemize}
\item[---] $\NDV$ between about $1\times 10^{16}$ and about 
$3\times 10^{18}\ \COit\ molecules\cdot\rm cm^{-2}$, which corresponds to 
$\NDv$ between about $2\times 10^{14}$ and $5\times 10^{16}\ \cOit\ molecules\cdot
\rm cm^{-2}$,
\item[---] $\nH\gsim 3\times 10^3\, cm^{-3}$ over much of the clouds' areas, except 
for $\nH\simeq 2\times 10^2\, cm^{-3}$ in the clouds' peripheries,
\item[---] $\Tk$ between 10$\,$K and 40$\,$K, the latter temperature found near
the H$\,$II regions.
\end{itemize}
\citet{Sakamoto94} chose to do their LVG analysis with diagrams of constant\hfil\break
$\XDvdr$ instead of constant $\NDV$.  They considered only values of $1\times 
10^{-4}$ and $1\times 10^{-5}\unit (km\cdot s^{-1}\cdot pc^{-1})^{-1}$.  Given
that the $\cOone$ clumps have velocity widths and sizes consistent with
$dv/dr$ of a $few\kms\cdot pc^{-1}$ \citep[see][]{Nagahama98}, the maximum 
$\XDvdr$ value should be a $few\times 10^{-5}\unit (km\cdot s^{-1}\cdot 
pc^{-1})^{-1}$, given the $\CO$ abundance mentioned in Appendix~A of Paper~I.  
Also, given that some structures have $dv/dr$ of around 100$\kms\cdot 
pc^{-1}$ --- even though seen with other tracers like CH$_3$OH \citep{Cernicharo99}
these structures are nonetheless real --- the minimum $\XDvdr$ should be 
around $10^{-6}\unit (km\cdot s^{-1}\cdot pc^{-1})^{-1}$.  Nevertheless, because such
structures have higher densities than those normally inferred from 
CO observations on parsec scales, the range of $\NDv$ values are probably
still roughly those of \citet{Sakamoto94}.  This range is larger than the range of 
$\nvtco$ values listed in Table~\ref{tbl-8}, but includes the range listed in that 
table.  The upper limits of $\NDv$ for \citet{Sakamoto94} and for Table~\ref{tbl-8} 
are similar to within a factor of 3, but the lower limits differ by more than an 
order of magnitude.  This disagreement is probably because
\citet{Sakamoto94} did not use the large-scale $\NDV$ of the entire cloud
as a rough lower limit on that of the 
clumps.  Their densities are consistent with the lower limit
in Table~\ref{tbl-8} for component~1, except for their density
for the peripheries.  This density is consistent with the lower limit we found for
component~0.  Accordingly, these peripheral regions may represent a low-density
envelope surrounding the entire clouds and are seen in projection along the clouds'
edges. The clumps in this envelope may be, in fact, the component~0 of the current
paper.  

The comparison between the kinetic temperatures of 
\citet{Sakamoto94} and those of the current paper is complicated 
by the high optical depths of the $\CO$ lines observed by the former. 
The high optical depths imply that the inferred densities and column densities
per velocity interval are biased towards of the surfaces of
clumps.  This is especially true
for the inferred kinetic temperatures, because, in the optically thick case, 
these temperatures are more directly related to the observed line radiation 
temperatures than are the other physical parameters.  The high optical depths
then mean that the warmer component will dominate the $\CO$ emission in these
lines.  We then simplistically assume that the $\Tdc<20\,$K 
subsample is essentially component~0 and that the $\Tdc\geq 20\,$K subsample
is component~1.  (Note that this is not consistent with the choice of
component used to compare the $\NDv$ and $\nH$ results.  If these other
physical parameters had been compared with the component-0 results instead
of those of component~1, then the agreement would still have been reasonable,
given the loose restriction on the parameters.)  With this approximation,
the gas kinetic temperatures as sampled by the $\CO$ are approximately
the same as the 140$\um$/240$\um$ dust color temperature, $\Tdc$. 
\citet{Sakamoto94} found $\Tk=40\,$K near the H$\,$II regions and 10--20$\,$K
away from the H$\,$II regions.  The current paper finds $\Td=\Tk\simeq 25\,$K
near the H$\,$II regions and 15--20$\,$K away from those regions, a range of
values less extreme than, and inside of, that of \citet{Sakamoto94}.  The range
is less extreme in the latter case because the spatial resolution is worse
by factors of 6 to 7 than those of \citet{Sakamoto94}. 

In short, the physical conditions derived here basically agree with
the previous results of \citet{Sakamoto94}.

\subsection{Column Density Determinations and the Two-Component Models}
\label{ssec42a}

As found in Paper~I (and Paper~Ia), the column densities derived from the
$\cOone$ line agreed significantly (at a confidence level better than 99.9\%) 
better with those derived from the far-IR continuum for the two-component models 
than for the one-component models.  An important difference between the two
components is that component~0 has unvarying gas and dust temperatures from
sightline to sightline, whereas component~1 has these temperatures varying
spatially (while maintaining a constant dust-gas temperature difference). 
Thus each sightline has two gas temperatures (and two dust temperatures),
except for those few sightlines where the component-0 and component-1
temperatures were the same.  Hence, two temperatures along each sightline
was necessary for recovering reliable column densities. 

This result is similar to that found by \citet{Schnee06}.  They compared the
continuum-derived column densities with those derived from extinction for the
Perseus and Ophiuchus molecular clouds.  The scatter in those plots more or
less matched that in a simulated cloud that assumed isothermal dust on each
sightline.  They concluded, therefore, that deriving reliable column densities
requires assuming variations of temperature along each sightline.  This
supports the result of the current work that at least two temperatures are needed 
on each sightline for estimating reliable column densities.

\subsection{$\bf\Delta T=0$\label{ssec43}}

The result that the gas and dust temperatures are the same is unexpected,
both theoretically and observationally.  For example, the theoretical model 
of PDRs applied to the Orion Nebula and its associated molecular gas 
\citep{Tielens85, Tielens85a} predicts $\DT\simeq -70$ to $+20\unit K$ for
cloud depths for which the dominant form of carbon is CO (i.e. $\rm A_v\gsim
3\,$mag).  This model, however, does not apply to multi-parsec scales in the 
Orion clouds: it uses a far-UV radiation field strength of $G_o=10^5$ and
a density of $2.3\times 10^5\ H\ nuclei\cdot\rm cm^{-3}$; both are too
high for the molecular clouds on larger scales.  The far-UV radiation field
on such scales is roughly $G_o\sim few$ (see Figure~17 and Section~\ref{ssec41} 
of W96) and the density could be as low as $few\times 10^3\ H_2\unit cm^{-3}$
(see Table~1).  Consequently, the ``standard" PDR model adopted by 
\citet{Mochizuki00} --- which assumes $\nH\sim 10^3\,cm^{-3}$, at cloud 
depths for which H$_2$ is the dominant form of hydrogen (i.e. $\NH\gsim
10^{21}\, cm^{-2}$), and $G_o=10$ --- is more appropriate for comparison with 
the work done here and yields $\DT\simeq -13$ to $+5\unit K$.  Considering the 
uncertainty in the $\DT$ of the current work, these theoretical values 
are factors of about 5 to 10 too large.  Given that there are many heating and 
cooling mechanisms in PDRs \citep[see][for a comprehensive description]{Tielens85} 
with different dependences on the density and on the radiation field (as well as on 
other quantities), adjusting the theoretical expressions for these many mechanisms 
to give $\DT$ near zero would only yield the desired 
result for an improbably narrow range of physical conditions (i.e., narrow range 
of $G_o$ and $\nH$ values).  On the other hand, simply increasing the gas-dust
thermal coupling by the required factor will easily achieve the desired result.
This corresponds to the $\DGC$ function developed by \citet{Burke83} and
represents a cooling mechanism of the gas due to its collisions with the dust
(and, of course if $\DGC<0$, it represents a heating mechanism of the gas due
to those collisions).  Therefore, increasing $\DGC$ by factors of 5 to 10 can
explain the current observations.

A commonly used form of $\DGC$ is that of \citet{Goldsmith01} and is given
as equation~(\ref{apg10}) in Appendix~\ref{appg}.  \citet{Goldsmith01} used the
\citet{Burke83} expression and adopted certain parameter values, including a
grain size that was a little too large.  Also the \citet{Burke83} expression 
only assumes a single grain size, and not the range of grain sizes that 
exists in the ISM \citep[see for example][]{MRN, Desert90}.  When a reasonably
realistic range of grain sizes is considered, the $\DGC$ of \citet{Goldsmith01}
is increased by factors of 3 to 4 (see Appendix~\ref{appg}).  This is not quite the
factor of 5 to 10 desired, but additional increases are easily possible when
one considers grains with non-spherical shapes or with projections on their
surfaces.  As mentioned in Appendix~\ref{appg}, $\DGC$ is proportional to the ratio
of the grain geometric cross-section to the grain volume (assuming uniform
grain density).  If the grains are elongated, then it would be easy to
increase this ratio.  (In reality it is the cross-section averaged over all
viewing angles that is important here.  However, even modest elongations of
a factor of a few would still result in a larger average cross-section than
a sphere with the same volume.)  Alternatively, projections on the grain
surface could also increase this ratio, but, since the relevant area is the 
cross-section rather than the total surface area, these projections would
have to be large compared to the grain size.   In any event, achieving an
additional factor of 2 is possible.  This would mean that simple geometric
considerations could increase the commonly used form of $\DGC$ by factors
of 6 to 8.  Therefore, $\DGC$ can indeed be
larger than had been previously assumed and could possibly explain the 
$\DT\simeq 0$ result for the Orion clouds.

Observationally, $\DT\simeq 0$ is unexpected as well \citep[e.g.,][]{Wu89,
Mangum99, Lis01}.  As discussed in the introduction, the different temperature
and density sensitivities of the dust continuum emission and gas line emission 
can result in incorrect inferences of the relative dust and gas temperatures at 
each point along the line of sight; the continuum and line emission preferentially
trace different regions of the ISM within the same line of sight.  In addition, in
many cases the uncertainty in the dust temperature or in the gas temperature, or
both, is large enough that $\Td=\Tk$ cannot be ruled out.  In \citet{Wu89}, for example, 
the uncertainties in $\Td$ were usually $\pm 1\,$K or $\pm 2\,$K.  The uncertainties in 
$\Tk$ were not explicitly listed for some sources, but would be at least about 10\% 
due to the stated calibration uncertainty.  Even though they did observe two lines of 
CO --- $\Jone$ and $\Jtwo$ of $\CO$ ---  they did not use their ratio to estimate 
$\Tk$; as shown in Figure~1 in the introduction of Paper~I, this gives only 
a very {\it un}interesting lower limit on $\Tk$.  Instead, they used the peak 
radiation temperature of each line, Rayleigh-Jeans corrected and corrected for the 
cosmic background, to estimate $\Tk$.  They assumed the lines to be 
optically thick and thermalized.  However, they also implicitly assumed that the gas 
fills the beam at the peaks of the lines.  As we have found in the current paper, 
this is not necessarily true, and may not be true even with the superior angular 
resolution of the \citet{Wu89} observations (a factor of 30 to 60 smaller beam).  
In addition, they used the 60$\um$ and 100$\um$ observations of IRAS.  
As stated before, the 60$\um$ emission suffers the contamination of emission from 
stochastically heated dust grains \citep[e.g.,][W96]{Desert90}.  However, because 
the \citet{Wu89} observations are on angular scales of 1$'$ to 2$'$, the radiation 
fields on such small size scales could be large enough that the 60$\um$ emission 
largely comes from grains in thermal equilibrium (i.e., the grains that 
would be stochastically heated in a normal interstellar radiation field reach thermal 
equilibrium in a strong radiation field).  A crude extrapolation of the trend in the 
data in Figure~10a of W96 suggests that the column densities derived from 60 
and 100$\um$ data will agree with those derived from longer wavelengths for $\Td
\simeq 60\,$K. Since $\Td=18\,$K for the general ISRF 
\citep[see W96,][]{Desert90}, and given that the ISRF is proportional to $\Td^{\beta 
+ 4}$, then radiation fields of at least $G_o\sim few\times 10^2$ 
ensure that the 60$\um$ emission originates largely from grains in thermal equilibrium.   
One of the sources observed by \citet{Wu89}, B35, has a radiation field of $G_o\simeq
30$ \citep[][and references therein]{Wolfire89}, roughly an order of magnitude too 
low to exclude the likelihood of stochastically heated grains contributing to the 
60$\um$ emission.  Therefore, for the source B35, and probably a few others in their 
list, the derived dust temperature overestimates the dust temperature of the large 
thermal equilibrium grains.  In short, their estimates of $\Tk$ are lower limits and 
their $\Td$ values are likely to be overestimates (especially in the case of B35).  
Accordingly, their conclusion that the observations are consistent with $\Td>Tk$ 
for the majority of their sources {\it still does not exclude the possibility of\/} 
$\Td=\Tk$ for these same sources.

The same can be said for the observations of \citet{Mangum99}.  They observed lines
of formaldehyde, H$_2$CO, towards dense gas condensations in NGC$\,$2024 in the
Orion$\,$B cloud with angular resolutions of 12$''$, 19$''$, and 30$''$.  They found
$\Tk$ values from around 50$\,$K to around 250$\,$K.  They then compared the derived
gas $\Tk$ values with the dust temperatures of \citet{Mezger92}, derived from the ratio 
of the 870$\um$- to the 1300$\um$-continuum 
emission at resolutions comparable to those of the H$_2$CO line
observations (i.e., 24$''$ and 8$''$, respectively).  These continuum observations
imply $\Td=19\,$K.  While, at face value, a difference between a $\Tk$ of 250$\,$K
and a $\Td$ of 19$\,$K may seem substantial, the uncertainties in $\Tk$
\citep[see Table~4 of][]{Mangum99} suggest that $\Tk-\Td$ is significant at levels 
of only 2 to 4$\,\sigma$.  These {\it would be\/} satisfactory levels of significance,
except that the correct uncertainty in $\Tk-\Td$ must also include the uncertainty in
$\Td$ as well --- an uncertainty that was ignored entirely.  \citet{Mezger92} state
a 20\% uncertainty in their continuum fluxes, implying an uncertainty of about 30\%
in the 870$\um$ to 1300$\um$ continuum.  This
uncertainty implies that $\Td=19\,$K is consistent with $\Td=7\,$K to
$\infty$. (Note that, even if we optimistically assume the continuum intensity ratio 
uncertainty to be only 20\%, the upper limit on $\Td$ would still be
$\infty$.)  In other words, the continuum observations do {\it not\/} place {\it any\/}
upper limit on $\Td$.  In fact, the difference between 19$\,$K and 250$\,$K corresponds
to only a 16\% change in the 870$\um$/1300$\um$ intensity ratio --- a change of
$\sim$0.5-$\sigma$.  Therefore, again, the observations {\it do not exclude the
possibility of\/} $\Td=\Tk$. 

In contrast, observations of giant molecular cloud cores in the Galactic Center
by \citet{Lis01} seem to genuinely rule out equal dust and gas temperatures.
They use continuum observations at a number of wavelengths from 45 to 850$\um$
and observations of the molecular lines of H$_2$CO, CS and other molecules to 
determine reliable dust and gas temperatures.  They find two components of dust: a 
warm component with $\Td\simeq 35\,$K, which dominates for $\lambda\lsim 100\um$, 
and a cooler component with $\Td\simeq 18\,$K, which dominates for $\lambda\gsim 
100\um$.  (Note that they also estimated the radiation field strength to be 
$G_o\simeq 500$ to 1000.  Therefore, the shorter wavelengths can also 
give reliable dust temperatures in this case.)  The molecular gas temperatures are 
$\Tk\simeq 60$ to 90$\,$K, implying gas-dust temperature difference as high as about 
70$\,$K.  However, given the appreciable foreground 
emission towards the Galactic Center and the lack of velocity information from the
continuum observations, contamination of these continuum observations by such 
foreground emission cannot be ruled out.  This foreground emission 
would be from the dust in the Galactic disk and has a temperature of about 18$\,$K 
\citep{Sodroski94}.  Consequently, only the 35$\,$K dust might be 
directly associated with the observed cores in the Galactic Center.  The molecules 
observed have transitions with high critical densities ($n_{_{crit}}\gsim 10^4\ 
H_2\cdot cm^{-3}$) and it could be argued that the observed transitions are only 
sampling the densest portion of the molecular gas.  However, given that the bulk of 
the gas in the Galactic Center is high-density gas \citep{Bally87a}, the observed
transitions are probably sampling most of the molecular gas in the observed cloud 
core.  Therefore, the gas-dust temperature difference cannot be any smaller than
about 25$\,$K, but is probably smaller than 70$\,$K. 

Considering the observations, we can draw an important
conclusion about $\DT$: it seems that $\DT\simeq 0$ is {\it not\/} excluded for 
a weak ISRF of $G_o\lsim 10^2$ and that it {\it is\/} excluded for a strong ISRF.  
Another possibility is that, given that the Galactic Center represents a unique 
Galactic environment, finding $\DT\ne 0$ may have more to do with other physical 
conditions than simply the strength of the radiation field and the gas density.  
Nevertheless, for now, a good working assumption is that $\DT$ is indeed near 0 
for $G_o\lsim 10^2$ and that $\DT$ is quite different from that, i.e. $|\DT|\gsim 
25\,$K, for $G_o\gsim 10^3$.  If this assumption is correct, then $\DT\simeq 0$ on 
multi-parsec scales for most molecular gas in the Galaxy, and $\DT\ne 0$
on these scales in the Galactic Center or in regions with large-scale star formation.
This has a number of consequences, including the following:
\begin{enumerate}
\item Galactic-scale molecular gas temperatures are nearly double the temperatures
previously believed.  Applying corrections for the cosmic background and for the
Rayleigh-Jeans approximation, the peak radiation temperatures found for the $\COone$
line in large-scale surveys of the Galaxy \citep[e.g.,][]{Sanders85} suggest that
$\Tk\sim 10\,$K.  If $\DT$ is indeed close to zero, then the true $\Tk$ is close to
that of $\Td$ on large scales, which is $\Td\sim 20\,$K \citep{Sodroski94}.  As
discussed in Section~3 of Paper~I and illustrated in  Figures~11, 18, 
23, and 26 of Paper~I, the molecular gas sampled by the $\COone$ line does
not fill the beam within each velocity interval within the line profile, especially 
if the linear beam size at the source is parsecs. 
\item The gas not completely filling the beam in each velocity interval may better 
explain the X-factor.  The usual explanation given for the 
N(H$_2$)/I(CO) factor is some variation of that of \citet{Dickman86}, that 
molecular clouds are virialized and that the line width indicates 
cloud mass and, therefore, cloud column density.  In fact, in some cases the 
velocity-widths of a cloud are only weakly correlated with column densities \citep[e.g., 
see][]{Heyer96}.  Consequently, a better explanation springs from having filling 
factors less than unity.  This will be discussed in a future paper \citep{W05b}.
\item $\DT=0$ constrains proposed explanations of the dust grain 
alignment that has been observed in the ISM \citep{Hiltner49, Hall49}.  For example, the
Davis-Greenstein mechanism is the relaxation of paramagnetic grains spinning in a 
magnetic field \citep{Davis51}.  This relaxation mechanism requires that $\Td\ne\Tk$
\citep{Jones67}.  However, there are number of other possible mechanisms that could 
explain dust grain alignment that require {\it no\/} such difference in temperatures
\citep[e.g., see][and references therein]{Lazarian97, Abbas04}. 
\end{enumerate}

Therefore, having equal gas and dust temperatures has a number of interesting 
consequences that are not {\it necessarily\/} contradicted by theory or observations.

\subsection{Cold Gas/Dust\label{ssec44}}

The two-component masses compared with those for the one-component masses in 
Table~6 of Paper~I imply about 60\% more mass of gas and dust in 
the Galaxy than previous estimates suggest.  These estimates are on the order of $5\times 
10^9\unit M_\odot$ of gas (i.e. molecular and atomic) in the Galaxy \citep{Dame93, 
Sanders92}.  If the model results for Orion are taken at face value and if these 
results apply to other clouds throughout the Galaxy, then this total gas mass 
increases to about $8\times 10^9\unit M_\odot$.  This increase is due to
some positions in component~1 having temperatures below 10$\,$K and as low as about 3$\,$K,
nearly that of the cosmic background.  Such cold dust and gas emits only weakly per unit
mass, allowing much gas and dust to be ``hidden" for the observed 
brightness.  Indeed, Table~7 of Paper~I lists the cold gas mass and it is about
40\% of the total mass listed in Table~6 of Paper~I (adopting Case~4 as the more
realistic of the two listed in Table~7 of Paper~I).  Accordingly, the total gas mass 
is the warm gas mass increased by about 60\% to allow for the cold gas mass.  Such an 
increase, especially if it applies to the entire ISM of the 
Galaxy, is substantial and its validity must be examined carefully.

As mentioned in Section~3.6 of Paper~I, the existence of this cold dust and gas depends on 
the basic assumption used in the modeling.  Even if this assumption has
provided a good physical description for most of the points in 
the $\rd$ versus $\Tdc$ plots (e.g., see Figures~20 and 24 of Paper~I), it does 
not necessarily apply to {\it all\/} of the triangular cluster of points from $\Tdc
\simeq 15$ to 21$\,$K and $\rd\simeq 10$ to 70$\MJkk$, especially to those with $\rd\geq 
30\,\MJkk$.  Indeed, it is the $\rd\gsim 30\MJkk$, $\Tdc=18\,$K points that require
$\Td$ as low as 3$\,$K when the basic assumption applies.  If these high-$\rd$, 
$\Tdc=18\,$K points are explained by other means, then such cold gas and dust may not
be present.  In other words, let us abandon, for the moment, all models
that use any form of the basic assumption, at least for these points.  There then exist 
a number of possibilities for the high-$\rd$, $\Tdc=18\,$K points:
\begin{itemize}
\item The $\NDv$ or $\nH$ is different from those of the rest of the points.  Models with
$\NDv$ or $\nH$ that vary smoothly with $\Tdc$ were discussed in Subsection~\ref{sssec384}.  
Even though such models have more difficulties than models using the basic assumption, it does
{\it not} exclude this possibility.  The $\NDv$
value would be higher or the $\nH$ value would be lower.  As we saw in
Subsection~\ref{sssec384}, $\nH$ would be as low as about 10$\unit cm^{-3}$.  
Figure~29 of Paper~I shows that these points occur mostly on some edges of the
Orion clouds.  A lower density for these points is consistent with them being on the
cloud edges.  However, such low densities imply a peak $\Tr(\cOone)$
either barely as strong as, or as much as an order of magnitude weaker than, the 
observed $\cOone$ line strength.  The other possibility of higher $\NDv$ 
would indeed give strong $\cOone$ emission, but implies $\NDv$ on the cloud 
edges a factor of a few higher than that for the cloud central regions.  This is
possible because $\NDv$ is not equivalent to N(H$_2$), but still seems somewhat 
implausible.
\item The points with high $\rd$ values at $\Tdc\simeq 18\,$K have appreciable emission of
dust associated with atomic hydrogen.  Given that this gas is largely found at some
cloud edges, this explanation seems reasonable; the atomic hydrogen and its associated 
dust would be found on molecular cloud edges, thereby providing the shielding necessary for 
the existence of the molecular gas.  However, we examined this in 
Subsection~\ref{sssec382} and found, based on the current data, that such emission was not 
likely to be important.  The
H$\,$I-associated dust contributes negligibly to the observed $\rd$ values.
Nonetheless, as discussed in that section, the gas 
and dust of component~1 with temperatures between 3 and 5$\,$K might still be explained by
H$\,$I-associated dust. 
\item The $\DT$ is different for the high-$\rd$, $\Td=18\,$K points.  Figure~7 of Paper~I
illustrates that $\DT$ varying smoothly from 0 to less than $-16\,$K (e.g., $\simeq -20\,$K) 
could account for the vertical extent of the points at $\Td=18\,$K in the $\rd$ versus $\Td$
plot.  However, Figure~17 of W96 suggests that $G_o$ is only a few and this would
give $\DT$ less extreme than those of \citet{Mochizuki00}, rendering $\DT\simeq -16$ to 
$-20\,$K unlikely.  Figure~7 of Paper~I also suggests that $\DT=+14\,$K
for these points could also account for their vertical extent.  But again, this is too 
extreme for the given $G_o$.  Also, this $\DT$ combined with $\Td=18\,$K would 
still result in cold gas, $\Tk=4\,$K.  A distinct $\DT$ for these points is an
unlikely explanation for their high $\rd$ ratio. 
\item The optical depth of the $\cOone$ line is high for these points, while $\Tk$ is still 
well above 3$\,$K.  This allows $\rd$ to be high, while obviating the need for cold
gas.  Given that the models require component~1 to be cold for these points (i.e., $\Tk=3$
to 10$\,$K) and component~0 to be 18$\,$K, then the models
already require at least some of this gas to be optically thick in the $\cOone$
line and some to be optically thin.  So, if the observations imply a high $\cOone$
optical depth for all of the gas, then the models would be in error.  The
gas density is near or above the critical density of the $\cOone$ transition, so
the line is close to LTE.  Consequently, the $\cOone/\COone$ intensity ratio
is a good estimate of the optical depth of $\cOone$.  This ratio for these points is about
0.3 --- significantly less than unity.  Therefore, at least some gas is indeed 
optically thin in $\cOone$, in agreement with the models. 
\item A lower $\cO$ abundance or a higher dust-to-gas ratio for these
points could account for the high $\rd$ values.  Accordingly, X($\cO$) would be
factors of 2 to 3 lower or $x_{_d}$ would be similar factors higher.  There is no
observational evidence for large changes in the $\cO$ abundance or the dust-to-gas mass ratio. 
Nevertheless, the models of \citet{Warin96} suggest that selective photodissociation of $\cO$ 
reduces its abundance by factors of 2 or 3 near molecular cloud surfaces.  If the triangular
cluster of points in the $\rd$ versus $\Tdc$ plot were affected by this reduced X($\cO$), then
correcting to the ``normal'' abundance would bring these points down by factors of 2 to 3. 
This would increase the lower temperature limit of the cold gas.  The bend in the model curve
from vertical to horizontal occurs more or less at the bottom of the triangular cluster and more or
less for a component-1 temperature of 7$\,$K.  Consequently, correcting for a {\it possible\/}
reduced $\cO$ abundance {\it could possibly\/} bring the lower temperature limit up to slightly 
less than 7$\,$K.  
\item A similar possibility would be an emissivity enhancement of the dust: the dust mass 
absorption coefficient, i.e. $\kappa_\nu$ at 240$\um$, would be unusually large, by factors 
of 2 to 3, for these points.  For example, a resonance in the dust absorption spectrum at 
240$\um$ could produce an increase in $\kappa_\nu$ at 240$\um$.  However, an increase by 
the same factor would be needed at 140$\um$ in order to maintain $\Tdc$ at about 18$\,$K.  
This resonance feature would be at least 100$\um$ wide.  This is very unlikely.  
Instead of a resonance, dust with opacities 2 to 3 times higher than normal dust at 
all FIR wavelengths is sufficient.  \citet{Dwek04}, for example, discusses
dust grains with far-IR opacities orders of magnitudes higher than the classical
silicate, graphite grains.  Mixing a very small portion of such grains with classical
grains could give a mix with an effective far-IR opacity easily factors
of 2 to 3 higher than considered here.  However, the observations require the opacity
to increase as the $\rd$ value increases towards the top of the triangular cluster.
Increasing opacity at far-IR wavelengths increases cooling as well.
Consequently, there would be an obvious overall trend to lower and lower 140$\um$/240$\um$
color temperature as $\rd$ increased --- resulting in a triangular cluster whose peak 
would be noticeably skewed towards lower colour temperatures.  This is not observed.  
Increasing the far-IR emissivity while keeping the 140$\um$/240$\um$ color temperature
constant requires having another warmer component mixed in that dominates
the dust continuum emission.  But this is nothing more than the original two-component
models that have been used up to this point.   Also, grains with such high far-IR 
emissivities are more likely to be responsible for such low dust temperatures (i.e. about 3 
-- 5$\,$K) rather than rule them out. 
\item Yet another possibility is that points towards the peak of the triangular cluster
are dominated by a component with a large value of $\rd$ and $\Tdc$ around 18$\,$K.  In
terms of the two-component models discussed in the current work, this is equivalent to
increasing the parameter $c_0$ for the peak of the triangular cluster.  
\end{itemize}
It seems that at least some alternatives 
to cold dust and gas may exist.  Overall, however, the evidence is far from convincing. 
In addition, simply abandoning the basic assumption leads to ad hoc interpretations.  This
is probably unjustified given the success of the basic assumption, and its associated models, 
at accounting for the overall trend in the data; the basic assumption and the models should
{\it not\/} be so lightly discarded. 

Can we explain the data without such cold dust and gas and
still use the models and the basic assumption?  In other words, is there some unexplored 
region of parameter space that permits a higher lower limit on the dust and gas 
temperature of component~1?  The answer is yes, but the changes are {\it not\/} dramatic. 
The lower limit to $\Tkone$ and $\Tdo$ depends on the position of the ``component-0 point''.
That is, the physical parameters of component~0 are constant in every respect and therefore
represent a single point in the $\rd$ versus $\Tdc$ plot.  In contrast, the gas and dust
temperatures of component~1 vary spatially, while the other parameters are held constant,
and therefore the parameters of component~1 represent a locus of points in this plot.  The
position of the component-0 point is just above the vertical section of the plotted 
two-component model curve.  When the $\rd$ of the component-0 point is higher
than the apex of the triangular cluster, then the lower limit on $\Tdo$ is higher.
In fact, this lower limit is roughly given by $\Tdo$ on the model curve at the
apex position.  Hence, the relevant parameter space area is where the component-0 
point has higher $\rd$.  This is easily accomplished by reducing the $\cO$ 
abundance {\it of component~0 only\/} by a factor of 2.  Doing this, and refitting the 
two-component models, raises the component-0 point from $\rd\simeq 65$ to about 110$\MJkk$.  
However, simply keeping the $\cO$ abundance fixed at its usual value and restricting $c_o$ to 
be $\geq 1$ results in even higher $\rd$ for the component-0 point.  In spite of the 
more promising position of this point, $\Tdo$ at the apex of the triangular cluster 
changes only by about 0.5$\,$K.  If we truncate the model curve above the $\Tdo=5\,K$ point 
(i.e. remove the temperatures {\it lower\/} than this), and keeping $\DT\geq 0\,K$ 
to keep $\Tkone\geq 5\, K$ as well, then the $\chi_\nu^2$ increases from 5.6 to 8.1, an 
increase excluded by the F-test at a confidence level of more than 90\%.  Accordingly, raising 
the component-0 point to higher $\rd$ merely stretches the vertical section of the 
curve between $\Tdo=2.8\,K$ and $\Tdo\simeq 5\,K$.  Consequently, we only increase the lower 
limit on $\Tdo$ from about 3$\,$K to about 3.5$\,$K.  In short, there is no compelling evidence 
that rules out cold dust and gas. 

Previous evidence for cold dust or gas is not compelling either, but does nonetheless come 
from a wide variety of observations \citep[e.g.,][]{Reach95, Merluzzi94, Ristorcelli98}.  
In addition, there is evidence for gas that had been previously undetected \citep[e.g.,][]
{Reach98, Cuillandre01}.  \citet{Reach95} used {\it COBE}/{\it FIRAS\/} continuum data with 
observed wavelengths from 104$\um$ to 2$\,$mm to infer a widespread cold component with dust 
temperatures 4 to 7$\,$K.  This component is found at all Galactic latitudes from 
the Galactic plane to the Galactic poles.  \citet{Lagache98} re-examined the {\it FIRAS\/} data 
and concluded that the cold component of \citet{Reach95} was not needed; the 
coldest component necessary was at about 15$\,$K. \citet{Finkbeiner99}, in yet another 
examination of the FIRAS data, concluded that the colder component has a temperature of
about 9$\,$K.  \citet{Merluzzi94} found a cold component with a temperature 
of either 15$\,$K or 7$\,$K, depending on whether the spectral emissivity index, $\beta$, was 
1.1 or 2, respectively.  \citet{Ristorcelli98} armed with continuum observations in four 
wavelength bands (i.e., 180-240, 240-340, 340-560, 560-1050$\um$) discovered a ``cold condensation" 
close to the Orion~Nebula with a temperature of 12.5$\pm 3\,$K.  This is not as cold as 
the 3 to 10$\,$K material discussed here, but nevertheless shows that dust with temperatures 
significantly lower than the 18$\,$K expected for dust heated primarily by the general ISRF is 
possible.  In contrast to these previous papers, the current paper infers 
a cold component (i.e. T$\simeq$ 3 to 10$\,$K) with{\it out\/} benefit 
of long-wavelength (i.e. $\lambda\gsim 1\,$mm) {\it continuum\/} data.   The long-wavelength 
data used here is the 2.7$\,$mm $\cOone$ {\it spectral line\/} in emission. 

If we accept for the moment that a cold component with temperatures of 3 to 10$\,$K does 
indeed exist within the Orion clouds, then the obvious question is how can 
such cold gas and dust exist without being strongly affected by the general ISRF or 
local stars?  \citet{Reach95} discuss a number of possible explanations in the context of
cold dust throughout the Galaxy, especially in high-latitude clouds.  We revisit some of
the proposed explanations of \citet{Reach95}, but in the context of the Orion clouds:
\begin{itemize}
\item[] {\it Shielding from the Interstellar Radiation Field.\/}  As \citet{Reach95} 
point out, attenuating the heating rate by a factor of 10$^3$ requires a {\it minimum\/} 
absorption equivalent to $A_v=20\,mag$ \citep{Mathis83}.  Given that the radiation field
is proportional to $\Td^6$ and that $\Td=3$ to 10$\,$K is the
cold dust temperature range, then attenuation factors of roughly 30 to 5$\times 
10^4$ are necessary.  This then requires minimum absorptions equivalent to $A_v\simeq
few\, mag$ to $A_v$ considerably more than 20$\, mag$ \citep{Mathis83}.  These correspond
to column densities of a $few\times 10^{21}$ to more than $3\times 10^{22}\ H\ nuclei
\cdot$cm$^{-2}$.  Over much of their area, the Orion clouds have N(H) closer to the former
value.  Consequently, shielding might explain the $\Td=10\,K$ material, but is
unlikely to explain the really cold material ($\Td\sim 5\,K$).  Also, since most of this
cold material is on the cloud edges, shielding probably will not
account for the cold dust and gas suggested by the models. 
\item[] {\it Fractal Grains.\/}  As previously mentioned, grains with enhanced 
submillimeter/FIR emission relative to UV/visible absorption, like the iron needles discussed 
by \citet{Dwek04} or fractal grains \citep[see][and references therein]{Reach95}, can have 
temperatures near that of the cosmic background.  \citet{Reach95} state that 
fractal grains may have greatly reduced mass for the given absorption cross-section;
the 60\% upward correction to the single-component mass would be revised considerably 
downward.  In addition, fractal grains may also have a reduced volume
for the given geometric cross-section, permitting $\DT\simeq 0$, as discussed previously
(see also Appendix~\ref{appg}). 
\item[] {\it Very Large Grains.\/}  Dust grains larger than the FIR wavelengths
that they emit do so very efficiently and, therefore, cool very efficiently.  
\citet{Reach95} find that the size distribution of these grains must steepen beyond that 
of the power-law index of $-3.5$ of \citet{MRN} to explain their observations.  This
steepening prevents a dust-to-gas mass ratio much higher than observed;
the power-law index of $-3.5$ out to a maximum grain radius of $~100\um$ increases
the dust-to-gas mass ratio by more than an order of magnitude.  In addition,  
the current work requires the observed grains to have a large geometric 
cross-section to volume ratio, thereby ruling out such large grains being the bulk of the 
cold dust.
\item[] {\it Long-Wavelength Emissivity Enhancement.\/}  \citet{Reach95} suggest that an
enhancement of the continuum emissivity at $\lambda\simeq 800\um$ could explain their 
observations instead of some hypothetical cold dust.  Obviously, this explanation does not 
apply to the current paper because the models have been applied to 240$\um$ continuum data
and 2.7$\,$mm {\it spectral line\/} data.  If the cold material predicted by the models 
for the Orion clouds is indeed ubiquitous in the Galaxy, then the \citet{Reach95} proposed 
enhancement of the long-wavelength continuum emissivity cannot rule it out. 
\end{itemize}
Of the explanations given above, fractal grains may be the most feasible. 

Thus the observations neither completely exclude nor strongly support the existence of 
cold dust and gas (i.e. T$\simeq 3$ to 10$\,$K) in the Orion clouds nor a widespread 
presence in the Galaxy as a whole.  The least unlikely alternative to dust and gas at 
temperatures of 3--5$\,$K is probably that mentioned in Subsection~\ref{sssec382}: the 
additional emission of dust grains associated with atomic hydrogen.  Nevertheless, it 
seems likely that cold material with temperatures of 7 to 10$\,$K does exist in the Orion 
clouds.  If cold dust and gas exist in the ISM in general with temperatures of $\sim 3$ 
to 10$\,$K, then fractal dust grains or iron needles may be the most credible reason and 
the 60\% upward correction to the Galactic ISM's mass due to this cold material might be 
revised downward substantially.  This downward revision is also necessary if the material 
at temperatures of 3 to 7$\,$K is really an artefact due to the other effects just described.

\subsection{The Millimeter Continuum to $\cOonebf$ Ratio as a Temperature Diagnostic\label{ssec45}}

Given that we can now characterize the $\Ib/I(\cOone)$ ratio as a function of physical parameters 
like the temperature, continuum and $\cOone$ observations can constrain 
the temperature of the dust and molecular gas.  In particular, using only ground-based observations 
to achieve such a constraint is advantageous.  Consequently, we 
examine one particular representative case: using the I(1300$\um$)/I($\cOone$) ratio --- hereafter, 
$\rdl$ --- to estimate the dust temperature, $\Td$.

\begin{figure}
\epsscale{0.68}
\plotone{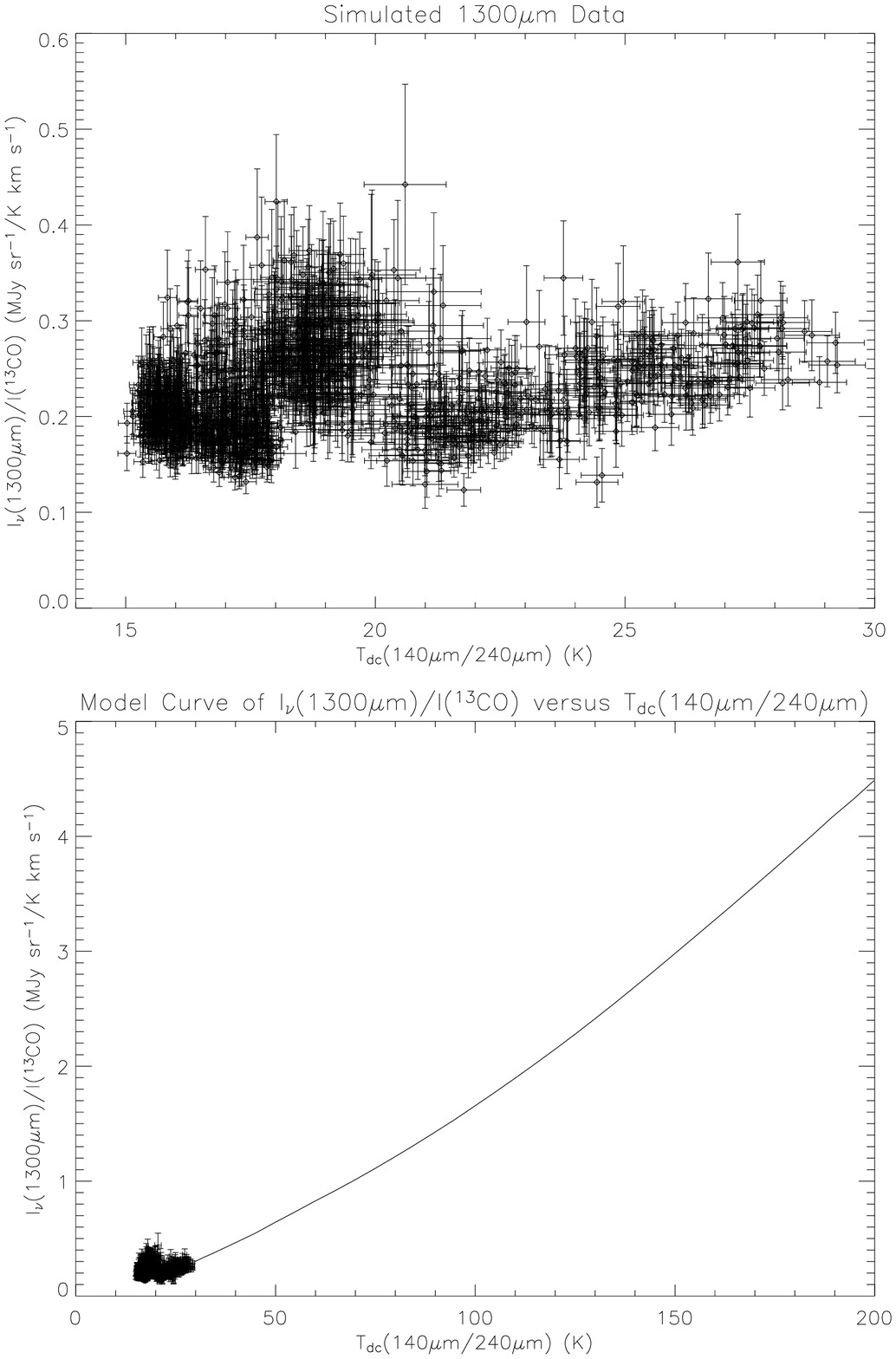}
\caption{These are plots of simulated 1300$\um$ continuum to $\cOone$ spectral line 
ratio data, i.e. $\rm I_\nu(1300\um)/\Ic$, versus simulated 140$\um$/240$\um$ color 
temperature data, i.e. $\Tdc$.  The simulations use the parameters listed in the
second column of Table~1 of Paper~II and a mass absorption coefficient appropriate
for $\lambda=1300\um$ (see details in Section~\ref{ssec45}).  The upper panel is the plot of 
$\rm I_\nu(1300\um)/\Ic$ versus $\Tdc$ for the temperature range $\Tdc = 14$ to 30$\,$K.  
The lower panel plots the same quantities, but for the larger temperature range of 
$\Tdc = 0$ to 200$\,$K.  The lower panel also shows the model curve out to $\Tdc=200\,$K.
\label{fig59}}
\end{figure}

To this end, the simulations described previously in Paper~II were used to generate a
model 1300$\um$ continuum map.  The adopted mass absorption coefficient at 1300$\um$,
$\kappa_\nu(1300\um)$, was determined from the adopted $\kappa_\nu(100\um)$ of 40$\unit
cm^2\cdot g^{-1}$ and scaling by $\nu^\beta$ for $\beta=2$, yielding $\kappa_\nu(1300\um)
= 40\unit cm^2\cdot g^{-1}\times (100/1300)^2$ or $0.24\unit cm^2\cdot g^{-1}$.  To adopt
a reasonable noise level, we note that current bolometers can achieve an rms noise value
of 2$\MJsr$ for a 1-hour integration in a 15$''$ beam at wavelength of 450$\um$ (D.~Hughes,
priv. comm.).  Assuming a noise proportional to $\lambda^{-2}$ and, depending
on the integration time and spatial averaging of the observations, a noise level of
0.075$\MJsr$ at 1300$\um$ is possible.  Therefore, this rms noise level is adopted for the
simulated 1300$\um$ map.  The simulated 1300$\um$ continuum map was divided by the 
simulated $\cOone$ map to produce the $\rdl$ map.  Figure~\ref{fig59} shows the resultant
simulated $\rdl$ values plotted against the simulated $\Tdc$ values --- the 140$\um$/240$\um$
color temperature.  This color temperature is, of course, not directly relevant to 1300$\um$
observations, but is included in Figure~\ref{fig59} to permit easy
comparison with the standard $\rd$ versus $\Tdc$ plots used throughout the
current paper.

The upper panel of Figure~\ref{fig59} shows that $\rdl$ is {\it not\/} useful as a
temperature diagnostic for dust {\it color\/} temperatures of $\sim 15$ to 30$\,$K; given 
that the vertical spread of the data points is about the size of the vertical error 
bars, we cannot unambiguously associate each $\rdl$ with a single $\Tdc$.  The temperature
dependence is weak in this range, because both the continuum and line observations
are close to the Rayleigh-Jeans limit.  This implies that neither component's emission
will overwhelmingly dominate over the other's for a larger range of temperature differences 
between the two components.  In contrast, at $\lambda=240\um$ and in the $\sim$3 to 30$\,$K
range of dust {\it physical\/} temperatures, component~1 goes from being overwhelmingly 
dominated by to overwhelmingly dominating the component-0 emission.  At $\lambda=1.3\,$mm,
component-1 goes from being dominated by only factors of a few to dominating by only a 
factor of $\sim$2.  Accordingly, combined emission of the two components has a narrow range 
of brightnesses. 

In contrast, the lower panel of Figure~\ref{fig59} demonstates that $\rdl$ can indeed be
a temperature diagnostic.  This plot shows the model curve up to temperatures of
200$\,$K, and the simulated map data points in the lower left corner.  The
curve goes from nearly flat (i.e. slope near 0) for $\Tdc\simeq 15$ to 30$\,$K to a power
law with
\begin{equation}
\rdl\propto\Td^{1.4}
\label{d1}
\end{equation}
above a threshold of 
\begin{equation}
\rdl\gsim 0.5\MJkk\ \ and\ \ \Td\gsim 50\, K\qquad .
\label{d2}
\end{equation}
(Note that $\Td$ can serve in place of $\Tdc$ for high temperatures, because the effect
of component~0 is negligible in this limit.)  Therefore, $\rdl$ is a useful temperature
diagnostic for temperatures above about 50$\,$K or $\rdl$ above $\sim 0.5\MJkk$.  Also, 
above this threshold the temperature sensitivity increases with increasing temperature.
This is not quite as strong as the $\Td^2$ rise expected in the LTE, high-temperature limit,
but is definitely much better high-temperature sensitivity than for the ratio of two rotational
lines of CO or for the ratio of two continuum bands. 

Naturally, the real situation is considerably more complicated than simply reading the
temperature from this one curve that {\it may\/} apply {\it only\/} to the Orion clouds.
As we recall, there are a number of physical parameters that affect this curve: $\Tdz$, $\nvcz$,
$c_0$, $n_{c0}$, $\nvco$, $n_{c1}$, $\DT$, $\kappa_\nu$ (or $\beta$), $x_{_d}$, and 
N(H$\,$I)/N(H$_2$).  Before we examine how variations of these parameters effect the model
curve, the curve in Figure~\ref{fig59} is adopted as the nominal curve.  The $\Td$ inferred
from an observed $\rdl$ using this nominal curve we call the nominal $\Td$.  Then the question 
is how the true $\Td$ differs from the nominal $\Td$ due to physical parameters differing from their 
nominal values (see the ``Input Values" column of Table~1 of Paper~II).  A reasonable goal is 
a temperature estimate within a factor of 2 of the true temperature for those
``hot spots" where $\rdl$ is above the threshold value of 0.5$\MJkk$.  If we vary the component-0
parameter values (i.e. the first four parameters listed above), then the true $\Td$ stays
within a factor of 2 of the nominal $\Td$ in most cases.  Specifically, varying $\Tdz$ from values 
lower than nominal to double this, $c_0\nvcz$ from factors 10 or
more lower (since this represents the optically thin limit), and $n_{c0}$ from 10$\unit cm^{-3}$
to 10$^5\unit cm^{-3}$ and higher (since this represents the LTE limit) changes $\rdl$ by
only 5-7\%.  Equivalently, the true $\Td$ differs from the nominal value by 7-10\%, well
within the desired factor of 2.  There are two potential difficulties.  One is if $c_0\nvcz$ is
two orders of magnitude higher than the nominal value, then the true $\rdl$ is 80\% lower
than its nominal value.  In practice this would not be a problem because the observed $\rdl$ would
be low enough that this observed position would not be identified as a hot spot; there would be no
false positives in the search for hot spots.  Nevertheless, any false hot spot is easily
identifiable with supplemental observations of the $\Jone$ line of $\Co$ or $\CO$; the $\Co/\cO$
intensity ratio would identify the position as having high $\cOone$ opacity.  The $\cO/\CO$ 
intensity ratio could be used similarly, but $\Co/\cO$ would be more reliable given that the
corresponding abundance ratio is within an order of magnitude of unity.  Another potential difficulty
is that, even though varying $n_{c0}$ over many orders of magnitude has a negligible effect on
$\rdl$ for $\Tdc\geq 50\,$K, $n_{c0}$ as low as 10$\unit cm^{-3}$ can raise the spur of data
points at $\Tdc\simeq 18\,$K (see Figure~\ref{fig59}) by about 70\%.  Accordingly, the threshold
listed above, expressions~(\ref{d2}), is raised to 
\begin{equation}
\rdl\gsim 0.85\MJkk\ \ and\ \ \Td\gsim 60\, K\qquad .
\label{d3}
\end{equation}
Densities as low as 10$\unit cm^{-3}$ are probably unlikely, so threshold~(\ref{d3}) is probably 
unnecessary.  Nevertheless, it can be used if extra caution is desired. 

Varying the component-1 parameters $\nvco$ and $n_{c1}$ results in the true $\Td$ being up to a
factor of 5 smaller than the nominal $\Td$.  The $\nvco$ value was increased by an order of 
magnitude or decreased similarly (or more because this is the optically thin limit). 
At the same time, the $n_{c1}$ value was decreased to 100$\unit cm^{-3}$ (again assuming that 
densities as low as 10$\unit cm^{-3}$ are unlikely) and increased to 10$^5\unit cm^{-3}$ (or more 
because this is the LTE limit).  Over most of this parameter space the true $\Td$ stayed within a 
factor of 2 of the nominal $\Td$.  However, the combination of $n_{c1}$ as low as 100$\unit cm^{-3}$ 
and $\nvco$ an order of magnitude larger gives a true $\Td$ up to a factor of 5 smaller than the
nominal $\Td$.  This low temperature case can be identified by observing the $\cOtwo$ line.  In
the nominal case, the ratio $\rm I(\cOtwo)/I(\cOone)$ varies between 1.9 and 3.4 for $\Tk = 50$ to 
200$\,$K.  In this high-$\nvco$, low-$n_{c1}$ case, $\rm I(\cOtwo)/I(\cOone)$ is 0.5 to 0.7.
Supplemental continuum observations help as well.  The $\rm I_\nu(450\um)/I_\nu(1300\um)$
ratio, for example, is sensitive to temperatures up to $\Td\simeq 60\,$K for a flux ratio 
uncertainty of 20\%.  Consequently, the observed $\rm I_\nu(450\um)/I_\nu(1300\um)$ ratio
places a lower limit on $\Td$.

Considering variations in $\DT$, the determination of $\Td$ is remarkably insensitive to such variations.  
For example, $\DT$ could range from $-80\,$K to $+80\,$K and the true $\Td$
is still within about 20\% of the nominal $\Td$.  Of course, the $\Tk$ would be quite different 
from $\Td$ in that case.  For $\DT=-80\,$K, the $\Tk$ is even higher than a nominal $\Td$ that 
was already high; this $\Tk$ is still within a factor of roughly 2 of the nominal $\Td$.   For 
$\DT=+80\,$K, $\Tk$ is quite small.  The $\rm I(\cOtwo)/I(\cOone)$ ratio is low in this case, 
but it is difficult to distinguish this case from the low-$\Td$, low-density case described in 
the previous paragraph using only this CO-line ratio.  Using the $\rm I_\nu(450\um)/I_\nu(1300\um)$ 
ratio in conjunction with the $\rm I(\cOtwo)/I(\cOone)$ may provide sufficient information to distinguish 
this high-$\Td$, low-$\Tk$ case from the low-$\Td$, low-density case. 

Dealing with anomalous values of the remaining parameters --- $\kappa_\nu$ (or $\beta$), $x_{_d}$, and 
N(H$\,$I)/N(H$_2$) --- is problematic but still possible.  If a source (a molecular cloud or part of
another galaxy) is mapped in $\rm I_\nu(1300\um)$ and $\cOone$, then the majority of the positions 
probably have roughly constant values for these parameters; the majority of these positions would
also have similar values of $\rdl$.  For example, it is easy to imagine that the molecular clouds in 
another galaxy have an unusually high or low dust-to-gas ratio, i.e. high or low $x_{_d}$, compared
with such clouds in our Galaxy.  Or we can imagine a cloud that has a non-negligible layer of atomic
gas (i.e. non-negligible N(H$\,$I)/N(H$_2$)).   In cases such as these, the $\rdl$
value itself would not be important, but that value in relation to the ``average" $\rdl$ values for the
observed source.  The hot spots are identified as those with $\rdl$ values large 
compared to the typical value for the source.  If, instead, such hot spots are not due to elevated 
temperatures but unusual $\kappa_\nu$, $x_{_d}$, and/or N(H$\,$I)/N(H$_2$) values, then such ``hot
spots" would {\it still\/} be interesting: they represent positions with unusual properties. 
Specifically identifying those unusual properties would involve the supplemental 
observations described above --- using the $\rm I(\cOtwo)/I(\cOone)$ and $\rm I_\nu(450\um)/I_\nu(1300\um)$ 
ratios --- and other observations when possible. 

Addressing the various points mentioned above, the following is a plausible observing plan:
\begin{enumerate}
\item First map the 1300$\um$ continuum and the $\cOone$ line.   The majority of points
establish the typical $\rdl$ value for the source.  Any points with high $\rdl$ values, above
some threshold that applies to the source, would be hot spots to
be ear-marked for further observations and study.  The $\rdl$ values for these hot spots will 
provide a temperature estimate to within a factor of 2 in the majority of cases. 
\item Map the $\cOtwo$ line.  The $\rm I(\cOtwo)/I(\cOone)$ ratio can help confirm whether the hot
spots are indeed hot.
\item Map the 450$\um$ (or shorter) continuum.  The $\rm I_\nu(450\um)/I_\nu(1300\um)$ ratio
is a further check on the temperature and can, in some cases, check whether $\DT = 0$ or not.
\item Map the $\Coone$ line.  The $\rm I(\Coone)/I(\cOone)$ ratio can confirm that $\cOone$
is indeed optically thin, thereby further checking the temperature inferred
from the $\rdl$ of the hot spots.  Even though the $\cOone$ is expected to be optically
thin on multi-parsec scales, as in the Orion clouds, verification of this can rule out other
possibilities.  
\end{enumerate}
All of these observations are possible from the ground.

Obviously, this is only one proposed observing plan out of many possibilities.  Similar plans
could be devised using other continuum wavelengths and other rotational lines.  Two such 
alternatives would be the $\rm I_\nu(2700\um)/I(\cOone)$ ratio or the \hfil\break
$\rm I_\nu(1300\um)/I(\cOtwo)$ 
ratio.  This is re-discovering the line-to-continuum (similarly, the continuum-to-line) 
ratio or the equivalent width, but for millimeter- or \hfil\break
submillimeter-wave molecular lines. 
Such equivalent widths are commonly used at visible, infrared, and centimeter wavelengths for 
inferring the physical properties of H$\,$II regions \citep{Osterbrock89, Spitzer78}.  Likewise, 
the equivalent widths of millimeter-wave molecular lines can provide important physical insights 
into molecular clouds (i.e. H$_{\it 2}$ regions). 

Nevertheless, the $\cOone$ line has a big advantage over higher rotational lines of $\cO$:
{\it relative\/} insensitivity to physical parameters like density and column density per
velocity interval.  This is why the $\rdl$ ratio can often predict $\Td$ to within a factor 
of 2, {\it despite varying $\nH$ and $\NDv$ by orders of magnitude.\/}  At the very least,
this method potentially places a realistic and interesting upper limit on dust and molecular 
gas temperatures.  

\section{Conclusions\label{sec5}}

Far-infrared continuum data from the {\it DIRBE\/} instrument aboard the {\it COBE\/} spacecraft were 
combined with $\cOone$ spectral line data from the Nagoya 4-m telescope to infer the large-scale 
(i.e. $\sim 5$ to $\sim 100\,$pc) physical conditions in the Orion molecular clouds.  The 
140$\um$/240$\um$ dust color temperatures, $\Tdc$, were compared with the 240$\um$/$\cOone$ 
intensity ratios, $\rd$, to constrain dust and molecular gas physical conditions.  In addition, 
such a comparison provides valuable insights into how the ratio of FIR/submillimeter/millimeter 
continuum to that of a $\cO$ (or $\Co$) rotational line can constrain temperature estimates of the 
dust and molecular gas.  For example, ratios of rotational lines or ratios of continuum emission 
in different wavelength bands often cannot place realistic upper limits on
gas or dust temperature, whereas the continuum-to-line ratio can place such limits. 

Two-component models fit the Orion data best.  One component has a fixed-temperature
and represents the gas and dust towards the surface of the clouds and is heated 
primarily by a kiloparsec-scale interstellar radiation field, referred to here as the general 
ISRF.  The other component has a spatially varying temperature and represents gas and 
dust towards the interior of the clouds that can be both shielded from the general ISRF and 
heated by local stars.  The model results and their implications are as follows:
\begin{itemize}
\item[1)] The inferred physical conditions are consistent with those derived from the 
large-scale observations of the $\Jtwo$ and $\Jone$ lines of $\CO$ by \citet{Sakamoto94}.
\item[2)] At least two gas (dust) temperatures are needed on the majority of sightlines
through molecular clouds for reliably estimating column densities.   This is supported
by the work of \citet{Schnee06}. 
\item[3)] The dust-gas temperature difference, $\Td-\Tk$ or $\DT$, is 0$\,$K to within 
1 or 2$\,$K.  If this result applies more generally to the Galactic-scale molecular ISM,
except for unusual regions such as the Galactic Center, then there are a number of 
implications:
\begin{itemize}
\item Dust-gas thermal coupling is factors of 5 to 10 stronger than has been previously
assumed.  Such factors may be due to the distribution of dust grain
sizes and grains with larger cross-section to volume ratios than that of
a simple sphere. 
\item Galactic-scale molecular gas temperatures are closer to 20$\,$K than to 10$\,$K,
because the emission from the CO rotational lines, even the optically thick $\COone$ line, 
does not fill the beam within the velocity interval about the line peak.
\item This CO emission that does not fill the beam provides a better explanation of the
N(H$_2$)/I(CO) conversion factor or X-factor.  Discussion of this is deferred to a later paper 
\citep{W05b}. 
\item Having $\DT$ nearly 0 constrains which mechanisms explain dust grain
alignment in the ISM.  A negligible dust-gas temperature difference rules out the 
Davis-Greenstein alignment mechanism, but not other possible mechanisms \citep[see][and
references therein]{Lazarian97, Abbas04}. 
\end{itemize}
\item[4)] Roughly 40--50\% of the ISM in Orion is cold (i.e. 10$\,$K) to very
cold (i.e. down to 3$\,$K) dust and gas.  Accordingly, there is roughly 60\% more gas
and dust in Orion than inferred from simple one-component models.  This {\it may\/}
also imply a similar increase in the estimated mass of entire Galactic ISM.  Fractal dust
grains \citep[see][and references therein]{Reach95} or iron needles \citep{Dwek04} may explain 
the low temperatures of this gas and dust  and, at the same time, may account for the high 
dust-gas thermal coupling needed to explain $\DT\simeq 0\,$K.  Nevertheless, alternative 
explanations that do {\it not\/} require cold dust and gas cannot be ruled out; the least 
unlikely of these other explanations is a contribution to the 240$\um$ continuum emission 
of the dust associated with atomic hydrogen.  The data suggest that the effect of the 
H$\,$I-associated dust is negligible, but still might permit raising the lower temperature 
limit of this cold gas and dust from 3 to 5$\,$K.  
\end{itemize}

The model parameter values derived from the fits to the $\rd$ versus $\Tdc$ plots
were used to create simulated 1300$\um$ continuum and $\cOone$ line maps.  These simulated
maps tested whether the millimeter continuum to $\cOone$ line intensity
ratio could constrain temperature estimates of the dust and molecular gas. The ratio 
$\rm I_\nu(1300\um)/I(\cOone)$, or $\rdl$, was found to estimate the dust
temperature to within a factor of 2 in most cases, provided that $\rdl$ was higher than 
a threshold level of 0.5$\MJkk$.  Supplemental observations of the $\cOtwo$ line and shorter
wavelength continuum would confirm the high temperatures in these 
high-$\rdl$ ``hot spots".   The results here can be easily generalized to other continuum
wavelengths and other rotational lines, even permitting interpretation of millimeter and
submillimeter molecular line {\it equivalent widths.\/}  And this is entirely possible with
only ground-based observations. 

The full potential of using millimeter continuum and $\cO$ (or $\Co$) rotational line
comparisons has yet to be realized.






\acknowledgments This work was supported by CONACyT grants \#211290-5-0008PE and 
\#202-PY.44676 to W.~F.~W. at {\it INAOE.\/} I am very grateful to W.~T.~Reach
for his comments and support.  I owe a great debt of thanks to Y.~Fukui and 
T.~Nagahama of Nagoya University for supplying the $\cO$ data that made 
this work possible. The author is grateful to R.~Maddalena and T.~Dame, who 
supplied the map of the peak $\COone$ line strengths and provided important 
calibration information.  I thank P.~F.~Goldsmith, D.~H.~Hughes, R.~Padman, 
W.~T.~Reach, Y. Fukui, M.~Greenberg, T.~A.~.D.~Paglione, G.~MacLeod, E.~Vazquez 
Semadeni, and others for stimulating and valuable discussions.



\appendix

\section{The Effect of Grain Size on Gas-Grain Thermal Coupling\label{appg}}

\citet{Burke83} describe the heat transfer rate from the gas to the dust
(or vice versa) as
\begin{equation}
\DGC = \nh\ngr\sgr\left({8k\Tk\over\pi\mh}\right)^{1\over 2}\acc (2k)(\Tk-\Td)\qquad ,
\label{apg1}
\end{equation}
where $\nh$ and $\ngr$ are the number densities of hydrogen molecules and dust grains,
respectively, $\sgr$ is the grain {\it geometric\/} cross-sectional area, $\Tk$ is the 
gas kinetic temperature, $\Td$ is the dust grain temperature, $\mh$ is the mass of the 
hydrogen molecule, and $\acc$ is the accommodation coefficient.  This coefficient is a 
measure of how well the temperature of the gas particles that have collided with the dust
grains accommodate to that of the grain surface \citep{Burke83}.  (Note that the $\nh$ 
and $\mh$ in this expression replace the $\rm n_{_H}$ and $\rm m_{_H}$ in 
expression~9 of \citet{Burke83} because the colliders considered here are hydrogen 
molecules.)  If we simplistically assume identical grains, then
\begin{eqnarray}
x_{_d} &=& {\ngr\mgr\over\nh\mh}
\label{apg2}\\
\noalign{\noindent or} 
\ngr &=& {x_{_d}\nh\mh\over\mgr}\qquad ,
\label{apg3}
\end{eqnarray}
where $\mgr$ is the mass of a single dust grain.  Assuming spherical grains
of uniform density, $\dgr$, yields
\begin{eqnarray}
\mgr &=& {4\over 3}\pi \agr^3\dgr
\label{apg4}\\
\noalign{\noindent and}
\sgr &=& \pi\agr^2\qquad ,
\label{apg5}
\end{eqnarray}
where $a$ is the grain radius.  Expression~(\ref{apg4}) substituted into (\ref{apg3})
gives
\begin{eqnarray}
\ngr &=& {3\over 4}\,{x_{_d}\nh\mh\over\pi\agr^3\dgr}
\label{apg6}\\
\noalign{\noindent and multiplying the above by (\ref{apg5}) yields}
\ngr\sgr &=& {3\over 4}\,{x_{_d}\nh\mh\over\agr\dgr}\qquad .
\label{apg7}
\end{eqnarray}
Note that expression~(\ref{apg7}) is nearly identical to expression~(5) of 
\citet{Goldsmith01}; he also included a $Q$-correction factor that is not necessary here 
because we are only concerned with the geometric cross-section and not the absorption
cross-section.  Note also that (\ref{apg7}) is proportional to the grain cross-section
to volume ratio.  (In reality, it is the grain cross-section to mass ratio that
is relevant, but, for a uniform grain density, this is equivalent to a proportionality
to the cross-section to volume ratio.)  Substituting (\ref{apg7}) into (\ref{apg1}) yields
\begin{eqnarray}
\DGC &=& {3\over 2}\,\left({8k^3\mh\over\pi}\right)^{1\over 2}\acc\,{x_{_d}\over
\agr\dgr}\,\nh^2\,(\Tk-\Td)\,\Tk^{^{1\over 2}}\qquad .
\label{apg8}\\
\noalign{\noindent Putting in the physical constants in cgs units gives}
\DGC &=& 7.11\times 10^{-36}\ \acc\,{x_{_d}\over\agr\dgr}\,\nh^2\,(\Tk-\Td)
\,\Tk^{^{1\over 2}}\qquad .
\label{apg9}
\end{eqnarray}
\citet{Goldsmith01} adopted the following values for the dust parameters:
$x_{_d}=0.01$, $\dgr=2\, g\cdot cm^{-3}$, and $a=1.7\times 10^{-5}\unit cm^{-3}$.
[Note that \citet{Goldsmith01} {\it apparently\/} adopted $a=1.7\times 10^{-7}\unit cm^{-3}$,
but this is near the lower limit of the dust grain size range \citep[see]
[]{Desert90}.  Also, and more importantly, that value of $a$ is inconsistent with
the numerical coefficients in the expressions for $\DGC$.  Finally, $a=1.7\times 10^{-5}\unit 
cm^{-3}$ was the actual {\it intended\/} dust grain radius (Goldsmith, {\it priv. comm.\/}).]
The recommended value of the accommodation coefficient is $\acc=0.3$ \citep{Burke83}.  
Consequently,
\begin{equation}
\DGC = 2.0\times 10^{-33}\ \nh^2(\Tk-\Td)
\,\left({\Tk\over 10\, K}\right)^{1\over 2}\qquad ,
\label{apg10}
\end{equation}
as per \citet{Goldsmith01}. 

However, as stated in Section~\ref{ssec43}, $\DGC$ must be factors of 5 to 10 larger to
explain the observations.  This is achieveable using a more realistic treatment of
grain sizes in deriving $\DGC$.  Specifically, a range of grain sizes must be considered 
instead of simplistically adopting a single radius.  For example, \citet{Desert90}
suggest that $a=15$ to 110$\,$nm for the big thermal equilibrium grains. 
Consequently, the 170$\,$nm adopted by \citet{Goldsmith01} is clearly too large; instead, 
some appropriately weighted mean of 15$\,$nm and 110$\,$nm is the most realistic choice for
the $a$ in expression~(\ref{apg9}).  From the work of \citet{MRN} we know that
\begin{eqnarray}
\ngr(\agr) &=& k_{_0}\agr^{-3.5}
\qquad ,
\label{apg11}\\
\noalign{\noindent where $\ngr(\agr)d\agr$ is the number density of grains with radii
between $\agr$ and $\agr+d\agr$ and $k_{_0}$ is a normalizing constant.  The $k_{_0}$ is 
found from}
\ngr(tot) &=& k_{_0}\int_\agi^\aga \agr^{-3.5}\, d\agr\qquad ,
\label{apg12}
\end{eqnarray}
with $\ngr(tot)$ as the number density of all the grains and corresponds to the $\ngr$ in 
(\ref{apg2}) and (\ref{apg3}).  From (\ref{apg12}), $k_{_0}$ can be expressed as
\begin{eqnarray}
k_{_0} &=& \ngr(tot)\ \hbox{$k_{_1}$}\qquad .
\label{apg13}\\
\noalign{\noindent $k_{_1}$ is another constant of the distribution and depends on 
$\agi$ and $\aga$.   Its exact form is not relevant to the derivation here, but is
nonetheless given below for completeness:}
k_{_1} &=& 2.5\ {(\agi\aga)^{2.5}\over a_{_{max}}^{2.5}- a_{_{min}}^{2.5}}\qquad .
\label{apg14}
\end{eqnarray}
To include the effect of the range of grain sizes on $\DGC$, the $\ngr\sgr$ in
equation~(\ref{apg1}) must be replaced with $\ngr(tot)\langle\sgr(\agr)\rangle_{_\agr}$,
where $\langle\sgr(\agr)\rangle_{_\agr}$ is the grain cross-section after averaging over
the size distribution.  Thus,
\begin{eqnarray}
\ngr(tot)\langle\sgr(\agr)\rangle_{_\agr} &=& \int_\agi^\aga \ngr(\agr)\,\sgr(\agr)
\, d\agr\qquad .
\label{apg15}\\
&=& 2\pi k_{_1}\,\ngr(tot)\,{\ax-\ai\over(\agi\aga)^{0.5}}\qquad ,
\label{apg16}
\end{eqnarray}
where equations~(\ref{apg11}), (\ref{apg5}), and (\ref{apg13}) were used.  The $\ngr(tot)$
on the right side must now be expressed in terms of the molecular gas density, $\nh$. 
This can be done using an expression analogous to that of (\ref{apg3}) that uses the 
range of dust sizes:
\begin{eqnarray}
\ngr(tot)\langle\mgr(\agr)\rangle_{_\agr} &=& x_{_d}\nh\mh\qquad ,
\label{apg17}\\
\noalign{\noindent or}
\int_\agi^\aga \ngr(\agr)\,\mgr(\agr)\, d\agr &=& x_{_d}\nh\mh\qquad .
\nonumber\\
\noalign{\noindent Applying equations~(\ref{apg4}), (\ref{apg11}), and (\ref{apg13}) 
yields}
{4\pi\over 3}\dgr\ngr(tot)\,\hbox{$k_{_1}$}\int_\agi^\aga \agr^{-0.5} d\agr 
&=& x_{_d}\nh\mh\qquad .
\nonumber
\end{eqnarray}
Integrating and solving for $\ngr(tot)$ gives us
\begin{eqnarray}
\ngr(tot) &=& {3\, x_{_d}\nh\mh\over 8\pi\dgr\hbox{$k_{_1}$}(\ax-\ai)}\qquad .
\label{apg18}\\
\noalign{\noindent Substituting (\ref{apg18}) into the right side of (\ref{apg16}) yields}
\ngr(tot)\langle\sgr(\agr)\rangle_{_\agr} &=& {3\over 4}\,{x_{_d}\nh\mh\over\aef\dgr}\qquad ,
\label{apg19}\\
\noalign{\noindent where}
\aef &\equiv& (\agi\aga)^{0.5}\qquad .
\label{apg20}
\end{eqnarray}
Equation~(\ref{apg19}) replaces the $\ngr\sgr$ that appears in (\ref{apg1}),
yielding an expression nearly identical to (\ref{apg8}) and (\ref{apg9}), except that
$a$ is replaced with $\aef$.   Therefore, the relevant grain radius in the 
expressions for $\DGC$ (e.g., equation~\ref{apg9}) is the geometric mean of the 
minimum and maximum grain sizes (i.e., equation~\ref{apg20}).  

The $\agi$ and $\aga$ should be those for the big
grains, rather than for the full range of dust sizes that also include the VSGs 
(very small grains) and the PAHs (polycyclic aromatic hydrocarbons) \citep[e.g., see]
[]{Desert90}.  The big grains are in thermal equilibrium and are the grains observed 
with the 140$\um$ and 240$\um$ {\it DIRBE\/} observations. 
According to \citet{Desert90}, the $\agi$ and $\aga$ are 15 and 110$\,$nm, respectively,
for the big grains.  This gives $\aef=41\,$nm and increases $\DGC$ by a factor of 4.2.
The full range of sizes over all grains, i.e. $\agi=0.4\,$nm and $\aga=110\,$nm,
results in $\aef=6.6\,$nm and $\DGC$ is increased by a factor of 26.  However,
there are at least two problems with using the full size range of grains.  One is that,
as mentioned above, only the big grains are relevant to the observations discussed here.
The second is that the treatment above implicitly assumes that, within the grain size
distribution, only the grain size changes; other grain properties, such as grain density 
and shape, are assumed constant despite changing 
grain size.  For example, going from (\ref{apg17}) to (\ref{apg18}) assumes 
that $\dgr$ is independent of $\agr$.  This is likely to be a bad approximation for
the full size range, especially when grain type varies with grain size \citep[e.g., 
see][]{Desert90}.  Therefore, the factor of 4.2 increase in $\DGC$ is appropriate when 
only considering the big grains.

A few corrections should be considered before using that factor of 4.2.  Given that only 
the big grains were used, we must correct for not using the full population of dust grains.  
Specifically, the dust-to-gas mass ratio used must be replaced by the mass ratio of the dust
in big grains to that of the gas.  The grain densities and sizes in \citet{Desert90}
imply that the big grains represent 76\% of the mass of the dust.  However, a more 
appropriate accommodation factor is necessary for the low dust and gas temperatures 
that considered here.  This suggests that $\acc$ is 0.4 instead of 
0.3 \citep[see][]{Burke83}.  Accordingly, increasing the accommodation factor
while decreasing the dust-to-mass ratio by similar amounts gives an overall 
correction of nearly unity (in fact it is about 0.9).  Another possible correction, or at 
least uncertainty, is the density adopted for the big grains.   While 
\citet{Goldsmith01} adopts $\dgr=2\unit g\cdot cm^{-3}$, \citet{Desert90} use $3\unit 
g\cdot cm^{-3}$ for the big grains.  This latter density brings the 4.2 factor down 
to about 3. 

In conclusion, considering a realistic range of grain sizes increases the gas-grain 
thermal coupling by factors of about 3 to 4.




\clearpage


\clearpage 

\clearpage 

\clearpage


\clearpage






\clearpage

\begin{deluxetable}{ccccccccc}
\tablecaption{Best-Fit Parameter Values for the Different Models\label{tbl-1}}
\tablewidth{0pt}
\tablehead{
\colhead{One-Component Models}\span\omit\span\omit\span\omit\span\omit
\span\omit\span\omit\span\omit\span\omit\\
\noalign{\bupskip}
}
\startdata
\undertext{LTE}\span\omit\span\omit\span\omit\span\omit
\span\omit\span\omit\span\omit\span\omit\\
\undertext{Full Sample}\span\omit\span\omit\span\omit\span\omit
\span\omit\span\omit\span\omit\span\omit\\
&&& $\DT$ & $\chi_\nu^2$ & $\nu$ &&&\\
&&& $-4$  & 16.5         & 673   &&&\\
\undertext{$\Td\ge 20\,$K}\span\omit\span\omit\span\omit\span\omit
\span\omit\span\omit\span\omit\span\omit\\
&&& $9$   & 9.0          & 140   &&&\\
\noalign{\bigskip}
\noalign{\smallskip}
\undertext{LVG}\span\omit\span\omit\span\omit\span\omit
\span\omit\span\omit\span\omit\span\omit\\
\undertext{Full Sample}\span\omit\span\omit\span\omit\span\omit
\span\omit\span\omit\span\omit\span\omit\\
&& $\DT$ & $\Ndv$            & $\nH$          & $\chi_\nu^2$ & $\nu$ &&\\
&& $-1$  & $3\times 10^{15}$ & $1\times 10^5$ & 16.9         & 671   &&\\
\noalign{\smallskip}
\undertext{$\Td\ge 20\,$K}\span\omit\span\omit\span\omit\span\omit
\span\omit\span\omit\span\omit\span\omit\\
&& $-3$  & $3\times 10^{15}$ & $6\times 10^3$ & 10.0         & 139   &&\\
\noalign{\bigskip}
\noalign{\hrule}
\noalign{\vskip 1pt}
\noalign{\hrule}
\noalign{\smallskip}
Two-Component Models\span\omit\span\omit\span\omit\span\omit
\span\omit\span\omit\span\omit\span\omit\\
\noalign{\smallskip}
\noalign{\hrule}
\noalign{\bigskip}
\undertext{Full Sample}\span\omit\span\omit\span\omit\span\omit
\span\omit\span\omit\span\omit\span\omit\\
$\DT$ & $c_0$ & $\Tdz$ & $\nvtcz$          & $n_{c0}$       & $\nvtco$            & $n_{c1}$       & $\chi_\nu^2$ & $\nu$ \\
0     & 0.04  & 18     & $5\times 10^{16}$ & $2\times 10^4$ & $8\times 10^{15}$   & $1\times 10^3$ &  5.7         & 667 \\
\noalign{\bigskip}
\noalign{\smallskip}
\undertext{Two Subsample}\span\omit\span\omit\span\omit\span\omit
\span\omit\span\omit\span\omit\span\omit\\
\undertext{$\Td < 20\,$K}\span\omit\span\omit\span\omit\span\omit
\span\omit\span\omit\span\omit\span\omit\\
$\DT$ & $c_0$ & $\Tdz$ & $\nvtcz$          & $n_{c0}$       & $\nvtco$            & $n_{c1}$       & $\chi_\nu^2$ & $\nu$ \\
0     & 1.0  & 18      & $5\times 10^{15}$ & $1\times 10^5$ & $2\times 10^{16}$   & $1\times 10^5$ &  4.6         & 525 \\
\noalign{\smallskip}
\undertext{$\Td\ge 20\,$K}\span\omit\span\omit\span\omit\span\omit
\span\omit\span\omit\span\omit\span\omit\\
0     & 0.4  & 18      & $5\times 10^{14}$ & $1\times 10^4$ & $5\times 10^{15}$   & $6\times 10^3$ &  8.2         & 135 \\
\enddata
			 
\tablecomments{$\DT$ and $\Tdz$ are in units of Kelvins.
The $\NDv$ quantities are in units of $\cOit\ molecules\cdot\ckms$.
The $n$ quantities are in units of $H_2\ molecules\cdot\rm cm^{-3}$.
All two-component models used the LVG code.
See Paper~I for discussion of the formal and systematic uncertainties.}

\end{deluxetable}

\clearpage

\begin{deluxetable}{cc}
\tablecaption{Best Estimates of Parameter Value Ranges\tablenotemark{a}
\label{tbl-8}}
\tablewidth{0pt}
\tablehead{
\colhead{Parameter} & \colhead{Range of Values}
}
\startdata
$\DT$\tablenotemark{b} & $-$1 to +2$\,$K \\
\noalign{\medskip}
$\Tdz$ & 16 to 19$\,$K\tablenotemark{c}\\
\noalign{\bigskip}
$c_0\nvtcz$ & $2.0\times 10^{14}$ to $5.0\times 10^{15}\ \cOit\ \ckms$\\
\noalign{\medskip}
$n_{c0}$ & $\gsim 20\ H_2\rm\ cm^{-3}$\\
\noalign{\bigskip}
$\nvtco$\tablenotemark{d} & $3\times 10^{15}$ to $2\times 10^{16}\ \cOit\ \ckms$ \\
\noalign{\medskip}
$n_{c1}$ & $\gsim few\times 10^3\ H_2\rm\ cm^{-3}$ \\
\enddata

\tablenotetext{a}{See Paper~II for details.}
\tablenotetext{b}{Assuming two-component models applied to {\it both\/} subsamples.}
\tablenotetext{c}{See Section~\ref{sssec386}.}
\tablenotetext{d}{For the two-component models applied to the two subsamples,
the $\nvtco$ value would be at the higher end of this range for the $\Tdc < 20\,$K
subsample and at the lower end for the $\Tdc\geq 20\,$K subsample.}


\end{deluxetable}

\clearpage


\begin{thebibliography}{}
\bibitem[Abbas et al.(2004)]{Abbas04} Abbas, M. M., Craven, P. D., Spann, J. F.,  Tankosic, D.,
	LeClair, A., Gallagher, D. L., West, E. A., Weingartner, J. C., Witherow, W. K., and
	Tielens, A. G. G. M. 2004, \apj, 614, 781
\bibitem[Bally et al.(1987)]{Bally87a} Bally, J., Stark, A. A., Wilson, R. W., and Henkel, C.
	1987, \apjs, 65, 13
\bibitem[Bally et al.(1991)]{Bally91} Bally, J., Langer, W. D., and Liu, W. 1991, \apj,
	383, 645
\bibitem[Boreiko and Betz(1989)]{Boreiko89} Boreiko, R. T. and Betz, A. L. 1989, 
	\apjl, 346, L97
\bibitem[Boulanger et al.(1998)]{Boulanger98} Boulanger, F., Bronfman, L., Dame, T. M., and
	Thaddeus, P. 1998, \aap, 332, 273
\bibitem[Boulanger et al.(1990)]{Boulanger90} Boulanger, F., Falgarone, E., Puget, J.-L., and
	Helou, G. 1990, \apj, 369, 136
\bibitem[Burke \& Hollenbach(1983)]{Burke83} Burke, J. R. and Hollenbach, D. J. 1983, \apj,
	265, 223
\bibitem[Cernicharo et al.(1999)]{Cernicharo99} Cernicharo, J., Pardo, J. R., González-Alfonso, E.,
	Serabyn, E., Phillips, T. G., Benford, D. J., and Mehringer, D. 1999, \apj, 520, L131
\bibitem[{\it COBE}/{\it DIRBE\/} Explanatory Supplement(1998)]{dirbex} {\it COBE} Diffuse Infrared Background 
	Experiment ({\it DIRBE\/}) Explanatory Supplement 1998,  version 2.3, ed. M. G. Hauser, T. Kelsall, 
	D. Leisawitz, and J. Weiland, {\it COBE} Ref. Pub. 98-A (Greenbelt, MD: NASA/GSFC), available in 
	electronic form from the NSSDC.
\bibitem[Cuillandre et al.(2001)]{Cuillandre01} Cuillandre, J.-C., Lequeux, J., Allen, R. J.,
	Mellier, Y., and Bertin, E. 2001, \apj, 554, 190
\bibitem[Dame(1993)]{Dame93} Dame, T. M. 1993, Back to the Galaxy, AIP Conf. 278, ed.
	S. S. Holt and F. Verter, New York : AIP, 267
\bibitem[Davis \& Greenstein(1951)]{Davis51} Davis, L., Jr. and Greenstein, J. L. 1951, \apj,
	114, 206
\bibitem[D\'esert et al.(1990)]{Desert90} D\'esert, F.-X., Boulanger, F., \& Puget, J. L.
	1990, \aap, 237, 215
\bibitem[Dickman(1975)]{Dickman75} Dickman, R. L. 1975, \apj, 202, 50
\bibitem[Dickman et al.(1986)]{Dickman86} Dickman, R. L., Snell, R. L., and Schloerb, F. P.
	1986, \apj, 309, 326
\bibitem[Dupac et al.(2000)]{Dupac01} Dupac, X., Giard, M., Bernard, J.-P., Lamarre, J.-M.,
	M\'eny, C., Pajot, F., Ristorcelli, I., Serra, G., and Torre, J.-P. 2000, \apj, 553, 604
\bibitem[Dwek(2004)]{Dwek04} Dwek, E. 2004, \apj, 607, 848
\bibitem[Finkbeiner et al.(1999)]{Finkbeiner99} Finkbeiner, D. P., Davis, M., and Schlegel, D. J.
	1999, \apj, 527, 867
\bibitem[Fixsen et al.(1999)]{Fixsen99} Fixsen, D. J., Bennett, C. L., and Mather, J. C. 1999,
	\apj, 526, 207
\bibitem[Goldsmith(2001)]{Goldsmith01} Goldsmith, P. F. 2001, ApJ, 557, 736
\bibitem[Goldsmith et al.(1997)]{Goldsmith97} Goldsmith, P. F., Bergin, E. A., and Lis, D. C.
	1997, \apj, 491, 615
\bibitem[Graf et al.(1993)]{Graf93} Graf, U. U., Eckart, A., Genzel, R., Harris, A. I., 
	Poglitsch, A., Russell, A. P. G., and Stutzki, J. 1993, \apj, 405, 249
\bibitem[Graf et al.(1990)]{Graf90} Graf, U. U., Genzel, R., Harris, A. I., Hills, R. E., 
	Russell, A. P. G., and Stutzki, J. 1990, \apjl, 358, L49
\bibitem[G\"usten et al.(1993)]{Gus93} G\"usten, R., Serabyn, E., Kasemann, C.,
	Schinkel, A., Schneider, G., Schulz, A., and Young, K. 1993, \apj, 402, 537
\bibitem[Hall(1949)]{Hall49} Hall, J. S. 1949, Science, 109, 166
\bibitem[Harris et al.(1991)]{Harris91} Harris, A. I., Hills, R. E., Stutzki, J., 
	Graf, U. U., Russell, A. P. G., and Genzel, R. 1991, \apjl, 382, L75
\bibitem[Harris et al.(1985)]{Harris85} Harris, A. I., Jaffe, D. T., Silber, M., and 
	Genzel, R. 1985, \apjl, 294, L93
\bibitem[Harrison et al.(1999)]{Harr99} Harrison, A., Henkel, C., and Russell, A. 1999,
	\apj, 303, 157
\bibitem[Heiles et al.(2000)]{Heiles00} Heiles, C., Haffner, L. M., Reynolds, R. J., and 
	Tufte, S. L. 2000, \apj, 536, 335
\bibitem[Heyer et al.(1996)]{Heyer96} Heyer, M. H., Carpenter, J. M., and Ladd, E. F. 1996,
	\apj, 463, 630
\bibitem[Hiltner(1949)]{Hiltner49} Hiltner, W. A. 1949, \apj, 109, 471
\bibitem[Howe et al.(1993)]{Howe93} Howe, J. E., Jaffe, D. T., Grossman, E. N., Wall, W. F., 
	Mangum, J. G., and Stacey, G. J. 1993, \apj, 410, 179
\bibitem[Johnstone \& Bally(1999)]{Johnstone99} Johnstone, D. and Bally, J. 1999, \apjl, 510, L49
\bibitem[Jones \& Spitzer(1967)]{Jones67} Jones, R. V. and Spitzer, L., Jr. 1967, \apj, 147, 943
\bibitem[Lagache et al.(1998)]{Lagache98} Lagache, G., Abergel, A., Boulanger, F., and Puget,
	J.-L. 1998, \aap, 333, 709
\bibitem[Lazarian et al.(1997)]{Lazarian97} Lazarian, A., Goodman, A. A., and Myers, P. C. 1997,
	\apj, 490, 273
\bibitem[Leisawitz \& Hauser(1988)]{Leisawitz88} Leisawitz, D. and Hauser, M. J. 1988, \apj,
	332, 954
\bibitem[Lis et al.(2001)]{Lis01} Lis, D. C., Serabyn, E., Zylka, R., and Li, Y. 2001, \apj, 
	550, 761
\bibitem[Mangum et al.(1999)]{Mangum99} Mangum, J. G., Wootten, A., and Barsony, M. 1999, \apj,
	526, 845
\bibitem[Mao et al.(2000)]{Mao00} Mao, R.Q., Henkel, C., Schulz, A., Zielinsky, M., 
	Mauersberger, R., St\"orzer, H., Wilson, T.L., and Gensheimer, P. 2000, \aa, 358, 433
\bibitem[Mathis et al.(1983)]{Mathis83} Mathis, J. S., Mezger, P. G., and Panagia, N. 1983
	\aap, 128, 212
\bibitem[Mathis et al.(1977)]{MRN} Mathis, J. S., Rumple, W., and Nordsieck, K. H. 1977, \apj,
	217, 425
\bibitem[Merluzzi et al.(1994)]{Merluzzi94} Merluzzi, P., Bussoletti, E., Dall'Oglio, G., and
	Piccorillo, L. 1994, \apj, 436, 286
\bibitem[Mezger et al.(1992)]{Mezger92} Mezger, P. G., Sievers, A. W., Haslam, C. G. T., 
	Kreysa, E., Lemke, R., Mauersberger, R., and Wilson, T. L. 1992, \aap, 256, 631
\bibitem[Mochizuki \& Nakagawa(2000)]{Mochizuki00} Mochizuki, K. and Nakagawa, T. 2000, \apj
	535, 118
\bibitem[Nagahama et al.(1998)]{Nagahama98} Nagahama, T., Mizuno, A., Ogawa, H., and Fukui, Y.
	1998, \aj, 116, 336
\bibitem[Osterbrock(1989)]{Osterbrock89} Osterbrock, D. E. 1989, Astrophysics of Gaseous Nebulae,
	Mill Valley : University Science Books
\bibitem[Plume et al.(2000)]{Plume00} Plume, R., Bensch, F., Howe, J. E., Ashby, M. L. N., 
	Bergin, E. A., Chin, G., Erickson, N. R., Goldsmith, P. F., Harwit, M., Kleiner, S.,
	Koch, D. G., Neufeld, D. A., Patten, B. M., Scheider, R., Snell, R. L., Stauffer, J. R.,
	Tolls, V., Wang, Z., Winnewisser, G., Zhang, Y. F., Reynolds, K., Joyce, R., 
	Tavoletti, C., Jack, G., Rodkey, C. J., and Melnick, G. J. 2000, \apjl, L133
\bibitem[Reach et al.(1995)]{Reach95} Reach, W. T., Dwek, E., Fixsen, D. J., Hewagama, T., 
	Mather, J. C., Shafer, R. A., Banday, A. J., Bennett, C. L., Cheng, E. S., Eplee, R. E., Jr.,
	Leisawitz, D., Lubin, P. M., Read, S. M., Rosen, L. P., Shuman, F. G. D., Smoot, G. F., 
	Sodroski, T. J., and Wright, E. L. 1995, \apj, 451, 188
\bibitem[Reach et al.(1998)]{Reach98} Reach, W. T., Wall, W. F., and Odegard, N. 1998, \apj,
	507, 507
\bibitem[Ristorcelli et al.(1998)]{Ristorcelli98} Ristorcelli, I., Serra, G., Lamarre, J. M., 
	Giard, M., Pajot, F., Bernard, J. P., Torre, J. P., De Luca, A., and Puget, J. L. 1998,
	\apj, 496, 267
\bibitem[Sakamoto et al.(1994)]{Sakamoto94} Sakamoto, S., Hayashi, M., Hasegawa, T., Handa, T.,
	and Oka, T. 1994, \apj, 425, 641
\bibitem[Sanders(1993)]{Sanders92} Sanders, D. B. 1993, Sky Surveys: Protostars to Protogalaxies,
	ed. B. T. Soifer, San Francisco : ASP 43, 65
\bibitem[Sanders et al.(1985)]{Sanders85} Sanders, D. B., Scoville, N. Z., and Solomon, P. M.
	1985, \apj, 289, 373
\bibitem[Schloerb et al.(1987)]{Schloerb87} Schloerb, F. P., Snell, R. L., and Schwartz, P. R. 1987,
	\apj, 319, 426
\bibitem[Schnee et al.(2006)]{Schnee06} Schnee, S., Bethell, T., Goodman, A. 2006, ApJ, 640, L47
\bibitem[Scoville \& Good(1989)]{Scoville89} Scoville, N. Z. and Good, J. C. 1989, \apj
	339, 149
\bibitem[Sellgren et al.(1990)]{Sellgren90} Sellgren, K., Luan, L., and Werner, M. W. 1990,
	\apj, 359, 384
\bibitem[Sodroski et al.(1994)]{Sodroski94} Sodroski, T. J., Bennett, C., Boggess, N., Dwek, E.,
	Franz, B. A., Hauser, M. G., Kelsall, T., Moseley, S. H., Odegard, N., Silverberg, R. F.,
	and Weiland, J. L. 1994, \apj, 428, 638
\bibitem[Sodroski et al.(1989)]{Sodroski89} Sodroski, T. J., Dwek, E., Hauser, M. G., and
	Kerr, F. J. 1989, \apj, 336, 762
\bibitem[Spitzer(1978)]{Spitzer78} Spitzer, L., Jr. 1978, Physical Processes in the Interstellar
	Medium, New York: Wiley
\bibitem[Swartz et al.(1989)]{Swartz89} Swartz, P. R., Snell, R. L., and Schloerb, F. P. 1989, \apj,
	336, 519
\bibitem[Tielens \& Hollenbach(1985)]{Tielens85} Tielens, A. G. G. M. and Hollenbach, D. 1985,
	\apj, 291, 722
\bibitem[Tielens \& Hollenbach(1985a)]{Tielens85a} Tielens, A. G. G. M. and Hollenbach, D. 1985,
	\apj, 291, 747
\bibitem[Wall(2007)]{W05} Wall, W. F. 2007, MNRAS, 375, 278 (see also astro-ph) (Paper~I)
\bibitem[Wall(2007a)]{W05a} Wall, W. F. 2007a, MNRAS, accepted (see also astro-ph) (Paper~II)
\bibitem[Wall(2007b)]{W05b} Wall, W. F. 2007b, MNRAS, submitted (see also astro-ph)
\bibitem[Wall et al.(1991)]{W91} Wall, W. F., Jaffe, D. T., Israel, F. P., and Bash, F. N.
	1991, \apj, 380, 384
\bibitem[Wall et al.(1996)]{W96} Wall, W. F., Reach, W. T., Hauser, M. G., Arendt, R. G., 
	Weiland, J. L., Berriman, G. B., Bennett, C. L., Dwek, E., Leisawitz, D., Mitra, P. M.,
	Odenwald, S. F., Sodroski, T. J., and Toller, G. N. 1996, \apj, 456, 566 (W96)
\bibitem[Warin et al.(1996)]{Warin96} Warin, S., Benayoun, J. J., and Viala, Y. P. 1996, \aap, 308, 535
\bibitem[Weiss et al.(2001)]{Weiss01} Weiss, A., Neininger, N., H\"uttemeister, S., and
	Klein, U. 2001, \aa, 365, 571
\bibitem[Werner et al.(1976)]{Werner76} Werner, M. W., Gatley, I., Harper, D. A., Becklin, E. E.,
	Loewenstein, R. F., Telesco, C. M., and Thronson, H. A. 1976, \apj, 204, 420
\bibitem[Wild et al.(1992)]{Wild92} Wild, W., Harris, A. I., Eckart, A., Genzel, R.,
	Graf, U. U., Jackson, J. M., Russell, A. P. G., and Stutzki, J. 1992, \aap,
	265, 447
\bibitem[Wilson et al.(2001)]{Wilson01} Wilson, T. L., Muders, D., Kramer, C., and Henkel, C.
	2001, \apj, 557, 240
\bibitem[Wolfire et al.(1989)]{Wolfire89} Wolfire, M. G., Hollenbach, D., and Tielens, A. G. G. M.
	1989, \apj, 344, 370
\bibitem[Wu \& Evans(1989)]{Wu89} Wu, Y. and Evans, N. J. 1989, \apj, 340, 307
\bibitem[Zhang et al.(1989)]{Zhang89} Zhang, C. Y., Laureijs, R. J., Chlewicki, G., Clark, F. O.,
	and Wesselius, P. R. 1989, \aap, 218, 231
\end{thebibliography}
\end{document}